\documentclass{elsart}
\usepackage{harvard}
\usepackage{graphicx}

\def\url#1{{\ttfamily\def\/{/\discretionary{}{}{}}#1}}

\begin{document} 
\begin{frontmatter}   
\title{TeV gamma rays from BL Lac Objects 
due to synchrotron radiation of extremely high energy protons}
\author{F. A. Aharonian}                                                       
\address{
MPI f\"ur Kernphysik, Saupfercheckweg 1,
D-69117 Heidelberg, Germany\thanksref{email}}
\thanks[email]{Felix.Aharonian@mpi-hd.mpg.de}
\begin{abstract}
One of  remarkable features of the gamma ray  blazar  Markarian 501 
is the reported shape of the TeV spectrum,  which during 
strong flares of the source remains essentially stable  despite dramatic 
variations of the absolute $\gamma$-ray flux.   
I argue that this unusual behavior  of the source could be explained 
assuming  that the TeV emission is a result of  synchrotron 
radiation  of  extremely high energy ($E \geq 10^{19} \, \rm eV$)  
protons   in highly magnetized ($B \sim 30-100  \, \rm G$) 
compact regions of the 
jet  with typical size   $R \sim 10^{15} - 10^{16}  \, \rm cm$ and  
Doppler factor $\delta_{\rm j} \simeq 10-30$. It is shown that
if protons are  accelerated at  the maximum possible 
rate, i.e. $t_{\rm acc}=\eta (r_{\rm g}/c)$ 
with so-called gyro-factor $\eta \sim 1$,   
the synchrotron cooling of protons 
could not only dominate  over  other radiative and non-radiative losses, 
but could also provide good fits (within uncertainties introduced 
by extragalactic $\gamma$-ray extinction)  to the $\gamma$-radiation  
of two firmly established TeV blazars - Markarian 501 and Markarian 421.  
Remarkably, if  the proton acceleration takes place in the regime dominated
by synchrotron losses, the spectral shape of the Doppler-boosted 
$\gamma$-radiation in the observer's frame is determined essentially by the 
self-regulated ``synchrotron cutoff'' at 
$\epsilon_0 \simeq  0.3  \ \delta_{\rm j} \eta^{-1} \ \rm TeV$.
The hypothesis of the proton-synchrotron origin of TeV  flares  
of BL Lac objects  inevitably implies  that the  energy  
contained in  the form of magnetic field  
in  the $\gamma$-ray emitting region  
exceeds the kinetic energy of accelerated  protons.
\end{abstract}
\end{frontmatter}

\section{Introduction}
Blazars are Active Galactic Nuclei (AGN) dominated by 
a highly variable component of non-thermal radiation 
produced in relativistic jets close to the line of sight
(e.g. Begelman et al. 1984,  Urry \& Padovani 1995). 
The dramatically  enhanced fluxes of the Doppler-boosted radiation,
coupled with the
fortuitous orientation of the jets towards the observer, make 
these objects  ideal laboratories to reveal the  underlying physics 
of AGN jets through multi-wavelength studies of 
temporal and spectral characteristics of  radiation from radio to 
very high energy  $\gamma$-rays (Ulrich et al. 1997) . 
First of all this concerns the BL Lacertae 
(BL Lac) objects - a
sub-population of blazars of which two prominent representatives,
Markarian 421 and Markarian 501, are firmly established as 
TeV $\gamma$-ray emitters. The flux variability 
on different time-scales  (and, plausibly, of different origin)
is a remarkable feature of TeV radiation of BL Lac objects. It
ranges from the spectacular 1996 May  15 flare  of Markarian 421 
with duration less than 1 h to the extraordinary high state of 
Markarian 501 in 1997 lasting several months  
(for review see Aharonian 1999; Catanese \& Weekes 1999).   
The recent multi-wavelength campaigns revealed
that the TeV flares of both objects, 
Markarian 501 (Pian et al. 1998; Catanese et al. 1997; 
Krawczynski et al. 2000; Sambruna et al. 2000)
and Markarian 421 (Buckley et al. 1996;
Maraschi et al. 1999; Takahashi et al. 1999), 
correlate with X-radiation on time-scales of hours or less.
This is often  interpreted as a strong argument in favor 
of the so-called synchrotron-Compton  jet emission models
in which the same population of ultra-relativistic  electrons
is responsible for production of both X-rays and TeV $\gamma$-rays
via synchrotron radiation and inverse Compton scattering, respectively  
(see, e.g., Ulrich et al. 1997). 
However, the very fact of correlation between X-ray and TeV $\gamma$-ray 
fluxes does  not yet rule out other possibilities, in particular 
the so-called hadronic models which assume that the observed 
$\gamma$-ray emission is initiated by accelerated protons
interacting with ambient gas or low-frequency radiation.

Generally, the hadronic models do not offer 
efficient $\gamma$-ray production mechanisms in the jet. 
For example, for any reasonable acceleration power of  protons, 
$L_{\rm p} \leq 10^{45} \, \rm erg/s$, 
the density of the thermal plasma in the jet 
should exceed  $10^{6} \, \rm cm^{-3}$ 
in order to interpret the reported TeV flares of Markarian 501 by 
$\pi^0$-decay $\gamma$-rays produced at $p$-$p$ interactions.
Therefore this mechanism could  be effectively realized only in 
a scenario like ``relativistic jet meets target'' 
(Morrison et al. 1984),  i.e.  assuming that $\gamma$-radiation 
is produced in dense gas  clouds that move across the jet 
(e.g. Dar \& Laor 1997)  
Recently a novel, ``non-acceleration''  version of 
$\pi^0$-decay $\gamma$-ray production by 
blazar jets was suggested  by Pohl \& Schlickeiser (2000).    
 
The {\em Proton Induced Cascade} (PIC) model 
(Mannheim 1993; Mannheim 1996) is  another attractive 
possibility for production of high energy $\gamma$-rays.   
This model relates the observed $\gamma$-radiation to the development 
of pair cascades in the jet triggered by secondary 
``photo-meson'' products  ($\gamma$-rays and electrons)  produced at 
interactions of accelerated protons with 
low-frequency synchrotron radiation.    
The efficiency of this model significantly  increases  
with energy of accelerated protons, therefore the 
postulation of an  existence of extremely high energy 
(EHE; $E \geq 10^{19} \, \rm eV$)  protons is a key 
assumption for  the PIC  model. In a  compact $\gamma$
production region of the jet with characteristic size less than 
$10^{16} \, \rm cm$, the protons could be accelerated to such high energies 
only in the presence of large magnetic field, $B \gg 1 \, \rm G$.
Even so, below I will show that the very fact of
observations of multi-TeV $\gamma$-rays from 
Markarian~421 and Markarian~501 allow a rather robust lower limit
on the ``photo-meson'' cooling time of protons, 
$t_{\rm p \gamma} \geq 10^7 \, \rm s$. 
This implies uncomfortably high  luminosity in EHE protons
which would be required to match  the observed TeV $\gamma$-ray 
fluxes. 

Meanwhile, at such conditions the synchrotron radiation of 
the EHE protons becomes a very effective channel of production
of high energy $\gamma$-rays.
In this paper I show that for a reasonable 
set of parameters, which characterize  the small-scale (sub-parsec) 
jets in Markarian 421 and Markarian 501, the synchrotron radiation 
of EHE protons not only may dominate over other possible 
radiative and non-radiative losses, but also could provide adequate 
fits to the observed TeV spectra of both objects\footnote{The 
synchrotron radiation of protons was actually 
included in the overall PIC code of Mannheim (1993), but  
the  effect of this process was somehow disregarded. 
The importance of the proton synchrotron radiation
in the PIC scenario was recently recognized, 
independent of the present paper, by M\"ucke \&
Protheroe (2000).}.
Moreover, this 
hypothesis could naturally  explain one of the remarkable 
features of TeV flares of  Markarian 501  - its essentially stable
spectral shape despite  spectacular variations  of the absolute TeV flux 
up to factor of 10 or more  on time-scales less than 1 day. 

\section{Synchrotron radiation of protons}
The comprehensively developed theory of synchrotron radiation of 
relativistic electrons (see, e.g.,  
Ginzburg \& Syrovatsky 1965; Blumenthal \& Gould 1970) 
can be readily applied to the proton-synchrotron radiation  
by re-scaling the Larmor frequency 
$\nu_{\rm L}=eB/2 \pi \, m c$  
by the factor $m_{\rm p}/m_{\rm e}\simeq  1836$. 
For the same energy of electrons and protons, $E_{\rm p}=E_{\rm e}=E$,
the  energy loss rate of protons  $({\rm d} E/{\rm d} t)_{\rm sy}$  appears 
$(m_{\rm p}/m_{\rm e})^4 \simeq 10^{13}$ times slower 
than the energy loss rate of electrons.  Also, 
the characteristic  frequency of the synchrotron radiation 
$\nu_{\rm c}=3/2 \ \nu_{\rm L} \, (E/m c^2)^2$  emitted by a
proton  is $(m_{\rm p}/m_{\rm e})^3 \simeq 6 \cdot 10^{9}$ times
smaller  than the characteristic frequency of synchrotron photons 
emitted by an electron of the same energy. 
Then the synchrotron cooling time of a proton, 
$t_{\rm sy} = E/({\rm d} E/{\rm d} t)_{\rm sy}$,
and the characteristic energy of synchrotron photons   
$\epsilon_{\rm c}=h \nu_{\rm c}$  produced 
in the magnetic field $B$ are
\begin{equation}
t_{\rm sy} =\frac
{6 \pi m_{\rm p}^4 c^3}
{\sigma_{\rm T} \ m_{\rm e}^2  \ E \ B^2}  
=4.5 \times 10^{4} \,  B_{100}^{-2} \ 
E_{\rm 19}^{-1} \; \rm s \, ,
\end{equation}
and
\begin{equation}
\epsilon_{\rm c}=
h \nu_{\rm c}=\sqrt{\frac{3}{2}} \
\frac{h e  B  E^2}{2 \pi m_{\rm p}^3c^5}
\simeq
87 \   B_{100} \ E_{\rm 19}^{2} \; \rm GeV \, ,
\end{equation}
where $B_{100}=B/100 \, \rm G$ and 
$E_{19}=E/10^{19} \, \rm eV$. Hereafter it is  assumed 
that the  magnetic field is distributed isotropically,
i.e. $B_{\perp}=\sin \psi \ B$ with $\sin \psi=\sqrt{2/3}$.

The average energy of synchrotron photons produced
by a particle of energy $E$ is equal to 
$\epsilon_{\rm m} \simeq 0.29 \ \epsilon_{\rm c}$ 
(Ginzburg \& Syrovatsky 1965). Correspondingly,  the characteristic 
time of radiation of a synchrotron photon of energy $\epsilon$ 
by a proton in the magnetic field $B$  is  
\begin{equation}
t_{\rm sy}(\epsilon) \simeq 2.2 \times 10^5 \ 
B_{100}^{-3/2} \ (\epsilon/1 \, \rm GeV)^{-1/2} \, \rm s \ .
\end{equation}
For comparison, the time needed for radiation of a synchrotron 
$\gamma$-ray photon  of  the same energy $\epsilon$ and in the same 
magnetic field,  but by an  electron,     
is shorter by a factor  $(m_{\rm p}/m_{\rm e})^{5/2} \simeq 1.5 \times 10^8$. 

\begin{figure}[htbp]
\begin{center}
\includegraphics[width=0.75\linewidth]{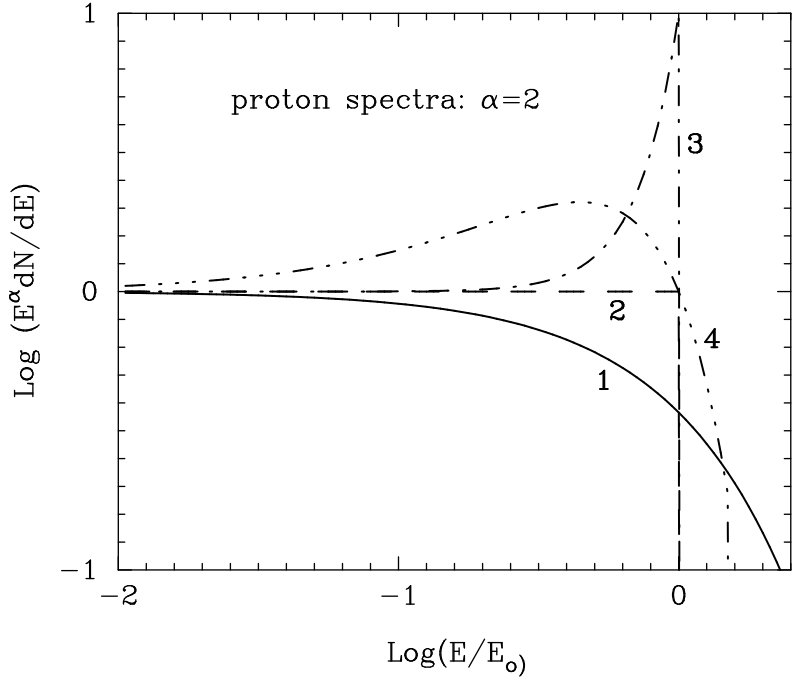}  
\caption{Different possible spectra of accelerated protons. 
At energies $E \ll E_0$ all spectra have power-law 
behavior with $\alpha_{\rm p}=2$, but in the region of the
cutoff $E_0$ they have essentially different shapes (see the text).}
\end{center}
\end{figure} 

The spectral distribution of synchrotron 
radiation emitted by a proton of energy E
is described by the equation 
\begin{equation}
P(E,\epsilon)=
\frac{\sqrt{2}}{h} \ \frac{e^3 B}{m c^2} \ F(x) \, ,
\end{equation}
where $x=\epsilon/\epsilon_{\rm c}$, and   
$F(x)=\ x \int_{x}^{\infty} {\rm d} x K_{5/3}(x) \,$; 
$K_{5/3}(x)$ is the modified Bessel function of 5/3 order. 
The function $F(x)$ could be presented in a simple analytical form 
(e.g. Melrose 1980):
\begin{equation} 
F(x)=C \ x^{1/3} \ \exp{(-x)} \, .
\end{equation} 
Numerical calculations show that with  
$C \approx 1.85$  this approximation provides 
very good, less than 1 per cent
accuracy in the region of the maximum at   $x \sim 0.3$,
and still reasonable (less than several per cent) accuracy in 
the broad  dynamical region   $0.1 \leq x \leq 10$.  

For the given energy spectrum $N_{\rm p}(E)$
of relativistic protons,  distributed isotropically in 
a source at a distance $d$ from the observer,
the differential flux of synchrotron radiation 
is defined as  
\begin{equation}
J(\epsilon)=\frac{\epsilon^{-1}}{4 \pi d^2} \ 
\int_{0}^{\infty} {\rm d}E \ P(E,\epsilon)\ N_{\rm p}(E)E \, .   
\end{equation}

Figure~1 shows four examples  of possible proton spectra.
The curve 1 corresponds to the most ``standard''  assumption 
for the spectrum of accelerated (e.g. by shock waves) 
particles - power-law with an exponential cutoff at energy $E_0$:   
\begin{equation}
N_{\rm p}(E)=N_0 E^{-\alpha_{\rm p}} \, 
\exp(-E/E_0) \, .
\end{equation}

The curve 2 corresponds to a  less realistic, truncated  
power-law spectrum, i.e. $N_{\rm p}(E) \propto E^{-\alpha_{\rm p}}$ 
at $E \leq E_0$, and $N_{\rm p}(E)=0$ at $E \ge E_0$.  

While the cutoff energy $E_0$ in the spectrum of
accelerated particles could be estimated quite confidently 
from the balance between the particle 
acceleration and the energy loss rates, the shape of the 
resulting spectrum in the cutoff region depends
on several circumstances - the specific mechanisms 
of acceleration and energy dissipation, 
the diffusion coefficient,  etc. 
For example, the recently revived interest in 
the diffusive shock acceleration of 
electrons resulted  in predictions that in the shock acceleration scheme 
one may expect not only spectral cutoffs,
but perhaps also pronounced pile-ups preceding   
the cutoffs (Melrose \& Crouch 1997; 
Protheroe \& Stanev 1999; Drury et al. 1999).
Two such spectra  are shown in Fig.~1.
The curve 3 represents the extreme class of spectra 
containing  a sharp (with an amplitude of factor of 10)
spike at the very edge of the spectrum. The 
curve 4 corresponds to a smoother spectrum with a modest    
pile-up (or ``bump'')  and super-exponential   (but not yet abrupt) cutoff. 

\begin{figure}[htbp]
\begin{center}
\includegraphics[width=0.75\linewidth]{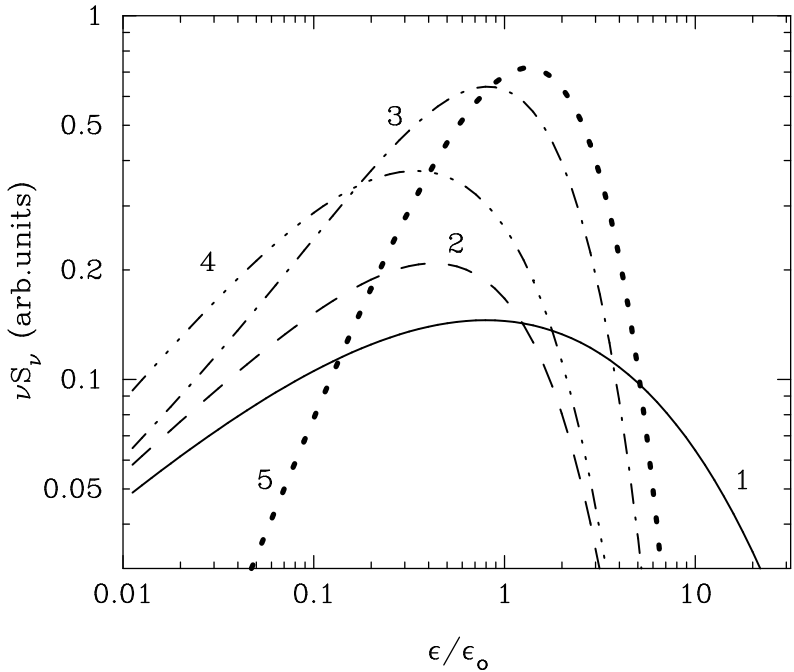}  
\caption{The Spectral Energy Distributions of the 
synchrotron radiation produced by protons with spectra shown in Fig.~1.
The curve 1 corresponds to the proton spectrum described by Eq.(7);
the curve 2 corresponds to the truncated proton spectrum;
the curve 3 corresponds to the proton spectrum with a sharp pile-up
and an abrupt cutoff at $E_0$;
the curve 4 corresponds to the proton spectrum with a smooth pile-up
and a super-exponential cutoff. For comparison, the SED of 
the  synchrotron radiation of mono-energetic protons, 
$xF(x)\propto x^{4/3} \exp{(-x)}$, is also shown (curve 5).}  
\end{center}
\end{figure}

The corresponding spectral energy distributions 
(SED: $\nu S_\nu=\epsilon^2 J(\epsilon)$) of 
synchrotron radiation are shown in Fig.~2. 
In the  high energy range, $\epsilon  \geq \epsilon_0$,
where $\epsilon_0$ is defined by Eq.(2) as  
$\epsilon_0=\epsilon_{\rm c}(E_0)$, 
the  radiation spectrum from the proton distribution 
with sharp pile-up and abrupt cut-off  
is quite similar to the synchrotron spectrum from mono-energetic 
protons, $x F(x)=x^{4/3} \ e^{-x}$; $x=\epsilon / \epsilon_0$.
This is explained by the radiation component associated with  the 
line-type feature  in the proton  spectrum at $E=E_0$.  
 
All synchrotron spectra shown in Fig.~2 exhibit, despite their 
essentially  different shapes,
cutoffs approximately  at  $x \sim 1$,   
if one defines the cutoff as {\it as  the energy at which the differential 
spectrum drops to $1/e$ of its extrapolated (from low energies) 
power-law value}. Therefore the energy  $\epsilon_{0}=\epsilon_{\rm c}(E_0)$ 
could be treated as an appropriate parameter representing the synchrotron cutoff 
for a quite broad class of proton distributions.
In the case of mono-energetic protons,
the cutoff energy coincides exactly  with $\epsilon_{0}$ 
(see Eq.~5). This is true also for the  power-law proton 
spectrum with exponential cutoff given by Eq.~(7), for 
which  the SED of the synchrotron radiation  has a shape close to 
$\nu S_\nu \propto \epsilon^{1/2} \ 
\exp{[-(\epsilon/\epsilon_0)^{1/2}]}$  (see below).

Below I will not specify the proton acceleration mechanisms, but
rather assume, with some exceptions notified otherwise,
that the accelerated protons in the source are 
represented  by a ``standard'' featureless spectrum given 
by Eq.~(7). For this  energy distribution of protons, 
the delta-functional approximation gives a simple analytical 
expression for the  differential spectrum of synchrotron radiation:
$J(E) \propto \epsilon^{-\Gamma} 
\exp{[-(\epsilon/\varsigma \epsilon_0)^{1/2}]}$, 
where $\Gamma=(\alpha_{\rm p}+1)/2$, and  $\varsigma$ is a 
free parameter introduced in order to adapt this  formula  to   
the accurate numerical calculations.

The spectra of synchrotron radiation
calculated in the delta-functional approximation for 
$\varsigma=1/3, \ 2/3, \ 1, \, \rm and \, 4/3$ 
(curves 2, 3, 4, and 5, respectively) are shown in Fig.~3,
together with the result of accurate numerical
calculations (curve 1).  It is seen that the best 
{\it broad-band} fit,  with accuracy better than $25 \%$ in the 
entire region up to $\epsilon \sim 20 \epsilon_0$,
is provided by $\varsigma=1$.   
Although $\varsigma=2/3$ gives, in fact,  better fit at
$\epsilon \leq \epsilon_0$, it significantly 
underestimates the flux above  $\epsilon_0$. And finally,
both $\varsigma=1/3$  and $\varsigma=4/3$ corresponding to 
the maximums in $S_\nu$ and $\nu S_\nu$ distributions of 
radiation emitted by a mono-energetic electron, do not 
adequately describe the radiation spectrum at 
$\epsilon \geq 0.1 \epsilon_0$. 
Thus, the most appropriate value for $\varsigma$ 
is  close to 1.  This proves that the synchrotron cutoff  
indeed takes place at $\epsilon_0$. 
At energies $\epsilon \ll \epsilon_0$ the radiation spectrum has a 
power-law shape with photon index 
$(\alpha_{\rm p}+1)/2$.  Note that 
the steepening of the $\gamma$-ray spectrum at higher energies 
is significantly  smoother ($\propto \exp{[-(\epsilon/\epsilon_0)^{1/2}]}$)
than the exponential cutoff in the spectrum of the parent  
protons\footnote{In fact, the numerical calculations show that 
the spectrum in the cutoff region drops even slower than 
$\exp{[-(\epsilon/\epsilon_0)^{1/2}]}$; compare the curves 1 and 
4 in Fig.~3.}. 
For the power-law proton distribution truncated at $E_0$, 
the synchrotron  spectrum beyond $\epsilon_0$ extends   
approximately as $\exp{[-\epsilon/\epsilon_0]}$). 

\begin{figure}[htbp]
\begin{center}
\includegraphics[width=0.75\linewidth]{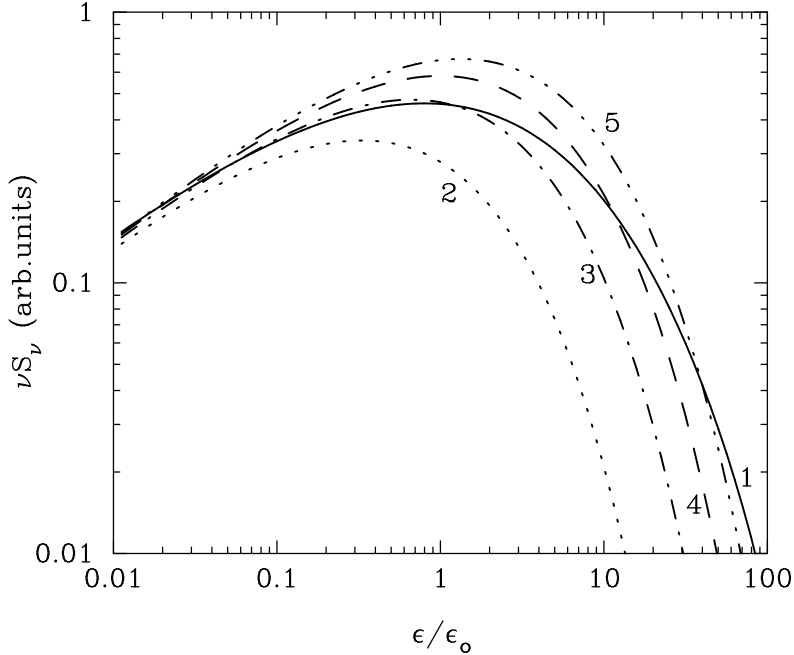}  
\caption{Spectral Energy Distribution (SED) of
the proton-synchrotron radiation calculated in the 
delta-functional approximation for the power-law 
spectrum of protons with a spectral index $\alpha_{\rm p}=2$ 
and an exponential cutoff at $E_0$ (Eq.~7). 
The curves 2, 3, 4, and 5 correspond to the 
parameter $\varsigma=1/3, 2/3, 1$ and 4/3, respectively.
The curve 1 correspond to accurate numerical calculations.} 
\end{center}
\end{figure}

Assuming now that the synchrotron radiation is emitted by a
relativistically moving cloud of plasma (often called
in literature as plasmon  or blob), in which
the spectrum of accelerated protons are represented by  
Eq.~(7),
the differential flux of radiation  
in the delta-functional approximation  
can be presented in the following simple form 
\begin{equation}
J(\epsilon)=A \epsilon^{-\Gamma} \, 
\exp[-(\epsilon/\epsilon_0)^{1/2}] \, ,
\end{equation}
where 
$A=(N_0 /48 \pi^2 d^2) \ (\sigma_{\rm T} 
B^2 m_{\rm e}^2/m_{\rm p}^4 c^3) \
a^{(\alpha_{\rm p}-3)/2} \delta_{\rm j}^{(\alpha_{\rm p}+5)/2}$,
$\Gamma=(\alpha_{\rm p}+1)/2$, 
$\epsilon_0= \delta_{\rm j} \epsilon_{\rm c}(E_0)$, and
$\delta_{\rm j}$ is the jet's Doppler factor.   

This convenient equation  could be used for the fit of 
TeV  observations from BL Lac objects 
in the framework of the proton-synchrotron
model, assuming that the spectrum of the accelerated protons
is given by Eq.~(7)

\section{Efficiency of proton synchrotron radiation}

The efficiency of a $\gamma$-ray production  mechanism 
is characterized by the
ratio of the radiative cooling time to the  
typical dynamical time-scale of the source 
corresponding to the minimum radiation variability in the jet's frame
\begin{equation}
t^\ast=\Delta t \  \delta_{\rm j} \simeq 1.08 \times 10^{5} \ 
\Delta t_{\rm 3h} \delta_{10} \: \rm s \, ,
\end{equation}
where $\Delta t_{\rm 3h}=\Delta t /3 \, \rm h$ is
the radiation variability in the observer's frame 
in units of 3 hour - a characteristic time-scale 
observed from Markarian 501 in TeV $\gamma$-rays  
(Aharonian et al. 1999a; Quinn et al. 1999), 
and $\delta_{10}=\delta_{\rm j}/10$.
The proton-synchrotron radiation 
becomes  an effective mechanism 
of $\gamma$-radiation with cooling time $\leq 10^5 \, \rm s$  
only when the photons are 
emitted by EHE protons with $E \geq 10^{19} \, \rm eV$
in a strong magnetic field close to 100 G. In this regime,
the synchrotron losses well dominate over 
non-radiative losses (see Fig.~4) 
caused by adiabatic expansion of the source or 
escape of particles from the source 
with characteristic time-scales $(c/v_0) \times t^\ast$ and
$(c/v_{\rm esc}) \times t^\ast$,  even in the case 
of relativistically expanding source ($v_0 \sim c$),
or energy-independent escape of particles with 
speed of light,  $v_{\rm esc} \sim c$.

It is interesting to compare the proton-synchrotron cooling 
time $t_{\rm sy}$ with  the photo-meson  cooling time, 
$t_{p \gamma}$. The energy  flux of low-frequency radiation 
that originates in the jet could 
be presented in the following form  
\begin{equation}
\nu S_\nu=10^{-12} \, g_{\rm fir} \,
(h \nu/0.01 \, \rm eV)^{-s+1} \; \rm erg/cm^2 s \, 
\end{equation}
where $g_{\rm fir}$ is a scaling factor indicating the 
level of the far infrared (FIR) flux at $h \nu=0.01 \, \rm eV$ 
($\lambda \sim 100 \, \mu \rm m$) in
unites of $10^{-12} \, \rm erg/cm^2 s$.  The flux 
normalization at FIR  wavelengths, 
which play  a major role in proton-photon interactions 
in the jet, makes the final results of calculations  
quite  insensitive to the choice of the spectral 
index $s$. In the case of Markarian 501, 
the power-law extrapolation of the observed synchrotron 
flux at soft X-rays ($\leq 1 \, \rm keV$) towards far infrared 
wavelengths with  $s=0.5$ gives  $g_{\rm fir} \simeq 0.3$.
On the other hand, assuming that the whole  radio-to-infrared 
flux observed from the direction of Markarian 501 
(see, e.g., Pian et al. 1998) is produced in a small-scale 
($\gamma$-ray emitting) 
jet, the  factor  $g_{\rm fir}$  could  be one order of magnitude larger. 

For  a source  with the {\em co-moving frame luminosity} 
$L^\prime(\nu^\prime) {\rm} \nu$,  
the  observer  at a distance  $d=c  z /H_0$ would 
detect a flux (Lind \& Blandford 1985)  
$S_\nu=\delta_{\rm j}^3 L^\prime(\nu/\delta_{\rm j})/4 \pi d^2$ 
(hereafter,  the value 
$H_0=60 \, \rm km/s \, Mpc$ will be used for the Hubble constant).  
The energy losses  of protons
in a low-frequency  photon field with a broad power-law
spectrum are dominated by photomeson-processes. For broad and  flat 
spectra of  target  photons  the photo-meson cooling time 
of protons can be estimated as 
$t_{p \gamma} (E) \simeq (c < \sigma_{\rm p \gamma}  
f >  n(\nu^\ast) h \nu^\ast )^{-1}$, 
thus for the flux given by Eq.~(10) with s=0.5, 
\begin{equation}
t_{p \gamma} 
\simeq 4.5  \times 10^{7} \ 
\Delta t_{\rm 3h}^2 \
\delta_{\rm 10}^{5.5}  \ (z/0.03)^{-2} \ 
g_{\rm FIR}^{-1} \ 
E_{19}^{-0.5} \, \rm s \, ,   
\end{equation}
where $<\sigma_{\rm p \gamma} f> \simeq 10^{-28} \, \rm cm^2$
is the photo-meson production cross section weighted by inelasticity
at the photon energy $\sim 300 \, \rm MeV$ in the proton rest frame,  
and $h \nu^\ast \simeq 0.03 \, E_{19}^{-1} \, \rm eV$ 
(see, e.g., Stecker 1968; M\"ucke et al. 1999).
Thus, for  both Markarian 501  and 
Markarian 421 with redshifts $z=0.034$ and $z=0.031$, respectively,
the photo-meson cooling time cannot be significantly less 
than $10^{7} \, \rm s$, unless
we assume a very high ambient photon density. Formally 
this could be  possible,  for example, adopting a much 
smaller blob size than it follows from the observed flux variability 
$\Delta t_{\rm 3h} \sim 1$, and/or  assuming a small Doppler factor
of the jet, $\delta_{\rm j} \ll 10$.
However, the photon density in the source cannot be 
{\em arbitrarily} increased, otherwise it would result in a catastrophic
absorption of TeV radiation inside the source. 

In the field of ambient photons   with differential power-law spectrum
$n(\nu) \propto  \  \nu^{-(s+1)}$,  
the optical depth for the photon-photon absorption is equal to   
$\tau_{\gamma \gamma}(\epsilon) =A(s) (\sigma_{\rm T}/2) 
\ h \nu_0 n(h \nu_0) R$,  
where $h \nu_0=4  (m_{\rm e} c^2)^2/\epsilon$, and
$A(s)=7/12  \cdot  4^{s+1} (s+1)^{-5/3}/(s+2)$ (Svensson 1987). 
For the spectral index $s=0.5$,  
\begin{equation}
\tau_{\gamma \gamma} \simeq 0.4 \ \Delta t_{\rm 3h}^{-1} \ 
\delta_{10}^{-5} \ g_{\rm fir} \  (z/0.03)^2 \
(\epsilon/1 \, \rm TeV)^{1/2} \, .
\end{equation}

The optical depth $\tau_{\gamma \gamma}$ 
increases with energy  as $\tau_{\gamma \gamma} \propto \epsilon^{1/2}$,
therefore  at 1 TeV it  should not  significantly exceed 1,  
otherwise the absorption of $\geq 10 \, \rm TeV$  $\gamma$-rays 
becomes unacceptably  large. For Markarian 501,  assuming 
that the low-frequency radiation  in the blob is 
described by Eq.~(10) with $s=0.5$ and adopting  a rather relaxed estimate 
for  $g_{\rm fir}  \sim 0.3$, 
the condition  $\tau_{\gamma \gamma}(1 \, \rm TeV) \leq 1$  
results in  a robust lower limit on the blob's Doppler factor 
$\delta_{\rm j} \geq  7$. 

\begin{figure}[htbp]
\begin{center}
\includegraphics[width=0.75\linewidth]{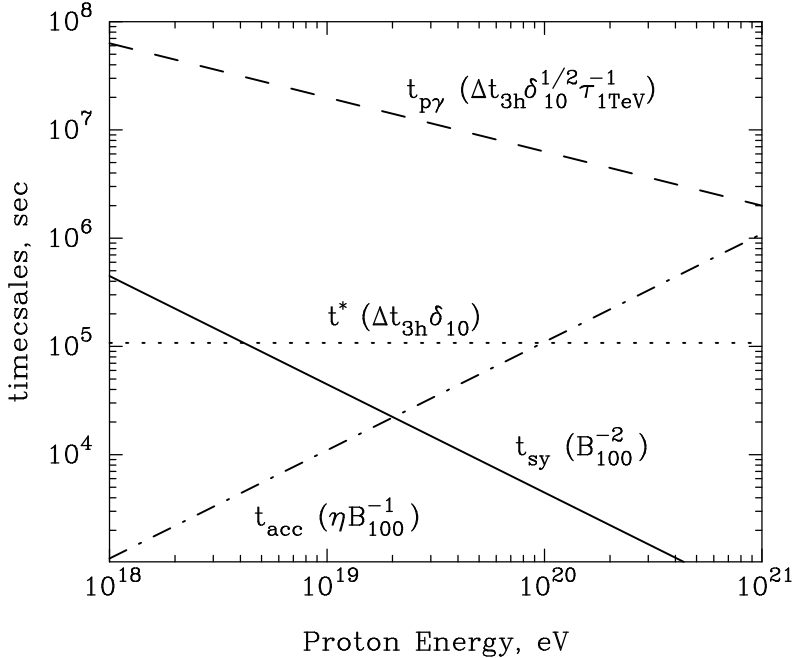}  
\caption{The characteristic acceleration and energy loss time-scales of protons
in the blob. At each curve the scaling factors (the products of relevant 
physical parameters) are  shown  (see the text).}
\end{center}
\end{figure}

It is worth noting that formally we may expect non-negligible flux
of TeV $\gamma$-rays  even at conditions when   $\tau_{\gamma \gamma} \gg 1$,
the radiation being contributed  from  the ``last layer'' 
of the  source with $\tau_{\gamma \gamma} \sim 1$.   
For a (quasi) homogeneous spherical source, this implies that 
the  ``TeV photosphere''  occupies  only  
$1/\tau_{\gamma \gamma}$ part of the total volume of the source.
On the other hand, the initial TeV radiation, produced throughout the 
entire source, is  re-radiated  in the form of secondary 
(cascade) electromagnetic radiation  in the energy band  below 
$\epsilon^\ast$, where $\epsilon^\ast$ is determined by  the condition 
$\tau_{\gamma \gamma}(\epsilon^\ast)=1$. Now assuming, for example, 
that at 1 TeV $\tau_{\gamma \gamma}=10$, 
the $\gamma$-ray luminosity of the source below 
$\epsilon^\ast=10 \, \rm GeV$ (for $s=0.5$)
should exceed the luminosity of the
``TeV photosphere'' by an order of magnitude.
This obviously contradicts the $\gamma$-ray observations
of Markarian 421, and especially Markarian 501. Therefore
we may conclude that the optical depth at 1 TeV,
$\tau_{\sf 1TeV}=\tau_{\gamma \gamma}(1 \, \rm TeV) \ll 10$.
            
From Eqs.(11) and (12) we  obtain a simple relation between
the photo-meson cooling time of protons, $t_{p \gamma}$, 
and the optical depth $\tau_{\sf 1TeV}$: 
\begin{equation}
t_{p \gamma} \simeq 1.8 \times 10^{7} \  
\Delta t_{\rm 3h} \ \delta_{\rm 10}^{1/2} \  
\tau_{\sf 1TeV}^{-1} \ 
E_{19}^{-1/2} \, \rm s.
\end{equation}

For a steeper  spectrum of low-frequency radiation,  e.g. a power-law with $s=1$,  
and for the same normalization to 
$\tau_{\sf 1TeV}$, the photo-meson cooling time is shorter:
\begin{equation}
t_{p \gamma} \simeq  10^{6} \  
\Delta t_{\rm 3h}  \, 
\tau_{\sf 1TeV}^{-1} \, 
E_{19}^{-1} \, \rm s .
\end{equation}

Even so,  $t_{\rm p \gamma}$ remains significantly larger 
than $t^\ast$, especially if we take into account that for 
$s=1$  the optical depth depends more strongly on energy,
$\tau_{\gamma \gamma} \propto \epsilon$, and therefore
the detection of TeV radiation from Markarian 501 beyond 10 TeV
is an indication  that  $\tau_{1 \sf TeV}$  should be 
significantly  less than   1.  

The source transparency  condition  for  multi-TeV $\gamma$-rays 
implies that in the TeV blazars the photo-meson processes 
proceed on significantly larger time-scales compared with 
$t^\ast$ and $t_{\rm sy}$ (see Fig.~4). In particular, 
for the power-law spectrum of low-frequency radiation 
with a (most likely)  slope $s=0.5$ 
\begin{equation}
\frac{t_{p \gamma}}{t^\ast} \simeq 1.7 \times 
10^2 \ \delta_{\rm 10}^{-0.5} \ 
\tau_{\sf 1TeV}^{-1}  \ E_{19}^{-0.5} \, ,
\end{equation} 
and 
\begin{equation}
\frac{t_{p \gamma}}{t_{\rm sy}} 
\simeq 4  \times 
10^2 \Delta t_{\rm 3h} \ \delta_{10}^{0.5} \
B_{100}^{-2} \ \tau_{\sf 1TeV}^{-1} \ E_{19}^{0.5} \, .
\end{equation} 

Thus, we conclude that 
for any reasonable assumption concerning the 
geometry of $\gamma$-ray production region, as well as 
the  spectral shape of low-frequency radiation in the blob, 
the detection of TeV  $\gamma$-rays from any  blazar would  
imply low efficiency of the photo-meson
processes in the jet,  unless the energy of protons does not 
significantly exceed  $10^{19} \, \rm eV$. In  compact 
$\gamma$-ray production region(s)  of  the jet with a typical size   
$R \leq 3 \times 10^{15} \Delta t_{\rm 3h} \delta_{\rm 10} \, \rm cm$,  
the proton acceleration to such high energies is possible only
in the  presence of magnetic field  
$B \gg 10 \, \rm G$ (see below).
At such conditions, however, the synchrotron radiation 
becomes  a more effective channel for conversion of the 
kinetic energy of  accelerated protons to very high energy $\gamma$-rays.
In principle, the difficulty with synchrotron losses could  be overcome by 
adopting  a  weak ($B \leq 1 \, \rm G$) magnetic field in the blob, 
but  assuming that the EHE protons are accelerated outside of the blob,
e.g. near the central compact object, and then 
transported along with the jet (Kazanas \& Mastichiadis 1999).  
However, this assumption  does not solve  the second problem 
connected with  the low efficiency of the   photo-meson processes
in the jet imposed by  the transparency condition for 
TeV $\gamma$-rays.
  
A regrettable  consequence  of this  conclusion is 
the suppressed TeV neutrino  flux.
However, it should be emphasized that this statement concerns {\it only}   
the objects seen  in TeV $\gamma$-rays. Meanwhile, the pair cascades 
initiated by  secondary electrons and $\gamma$-rays from $p \gamma$
interactions  may still remain 
a viable possibility for other AGN, in particular for 
the powerful GeV blazars detected by EGRET (Mukherjee et al. 1997),
as well as for radio-quiet AGN (e.g. Sikora et al.  1987),  
where the  radiation density is much higher than in the BL Lac objects, 
and, more importantly, the photo-meson cooling 
time of EHE protons is not constrained  by  the severe 
TeV $\gamma$-ray transparency  condition.

\section{Self regulated synchrotron cutoff}

The energy spectrum of synchrotron radiation  depends on the spectrum 
of accelerated protons and the jet's Doppler factor.  The
high energy cutoff in the spectrum of protons 
is determined by the balance between the particle 
acceleration  and cooling times. 
It is convenient to present the acceleration time 
of particles $t_{\rm acc}$ in the 
following general form 
\begin{equation}
t_{\rm acc}=\eta(E) \ r_{\rm g}/c = 1.36 \times 10^{4} \ E_{19} \ 
B_{100}^{-1} \  \eta(E) \; \rm s \, ,
\end{equation} 
where $r_{\rm g}=E/(e B_{\perp})$.
The so-called gyro-factor $\eta(E) \geq 1$ 
characterizes the energy-dependent rate of acceleration. 
For almost all proposed models
of particle acceleration in different astrophysical environments, 
$\eta$ remains a rather uncertain model parameter.
This is true especially for the small-scale jets 
of blazars, where the nature of the 
acceleration mechanism itself  remains highly unknown.  
On the other hand, any postulation of acceleration 
of EHE protons in compact $\gamma$-ray production 
regions of the 
small-scale jets actually  implies that the parameter $\eta$ 
should be close to 1, which corresponds 
(independent of a specific mechanisms of acceleration)
to the maximum (theoretically possible) acceleration rate 
based on a simple geometrical consideration (Hillas 1984).
In the case of {\it diffusive shock acceleration} in the blazar jets 
the parameter $\eta$ typically is expected  (Henri et al. 1999)
to be  larger than 10 (see, however, Bednarz \& Ostrowski 1996).  
Therefore,  perhaps  more effective acceleration  
mechanisms should be invoked for production of EHE protons 
in small-scale AGN jets.   
An interesting possibility could be the particle
acceleration   at  the annihilation of magnetic fields 
in the fronts of Poynting flux dominated jets 
(Blandford 1976; Lovelace 1976). It has been argued 
that this mechanism could provide 
effective  acceleration of EHE protons 
with  $\eta \sim 1$  (Haswell et al. 1992). 

The discussion  of particle acceleration mechanisms  
is beyond the framework of this paper, and therefore here 
the assumption  of $\eta \leq 10$  should be  treated just 
as an  working hypothesis.  If true, during the characteristic  
time $t^\ast$ given by Eq.~(9), the protons  could be accelerated
in the jet with $B \sim 100 \, \rm G$ and $\delta_{\rm j} \sim 10$ 
up to energies $E \sim 10^{20} \, \rm eV$.  At such conditions the losses 
of highest energy protons are dominated by the proton-synchrotron 
radiation, and therefore the cutoff energy $E_0$ is determined by the 
condition $t_{\rm sy}=t_{\rm acc}$ :
\begin{equation}
E_{0}=(3/2)^{3/4} \ 
\sqrt{\frac{1}{e^3 B}} 
\ m_{\rm p}^2c^4  
\simeq 1.8 \times 10^{19} \  B_{100}^{-1/2} \ \eta^{-1/2} \,  \rm eV \, .    
\end{equation}
Note that the relevant cutoff in the electron spectrum appears much earlier, 
$E_{e,0}=(m_{\rm e}/m_{\rm p})^2 E_{0} \simeq 5.3 \times 10^{12} 
\ B_{100}^{-1/2} \ \eta^{-1/2} \, \rm eV$.  

Substituting Eq.~(18) into Eq.~(2) we find that the position of the 
cutoff in the spectrum of  the proton-synchrotron radiation   
is determined by two fundamental physical
constants, the proton mass $m_{\rm p}c^2=938 \, \rm MeV$, 
and fine-structure constant $\alpha_{\rm f}=1/137$.   
It depends only on the   parameter $\eta$,  but not on the  magnetic field:
\begin{equation}
\epsilon_0 =\frac{9}{4} \ \alpha_{\rm f}^{-1} \ m_{\rm p} c^2 \ \eta^{-1} 
\simeq 0.3 \ \eta^{-1} \, \rm TeV \, .
\end{equation}    
For  the  electron-synchrotron radiation, 
the  universal cutoff  appears at 
$\epsilon^{\rm (e)}_0=(m_{\rm e}/m_{\rm p}) \  
\epsilon^{\rm (p)}_0 
=9/4 \ \alpha_{\rm f}^{-1} \ m_{\rm e} c^2 \ \eta^{-1} 
\simeq 0.160 \ \eta^{-1} \, \rm GeV$. 

Comparing now the minimum (energy-independent) 
escape time of protons 
$t_{\rm esc} \sim R/c \simeq 3.3 \times 10^4 \ R_{\rm 15} \xi \  \rm s$, 
with the synchrotron cooling time of the protons responsible for production of 
$\gamma$-rays in the cutoff region, 
$t_{\rm sy} \simeq 2.4  \times 
10^4 \ B_{100}^{-3/2}  \eta^{1/2}  \rm \ s$
we   find the  following condition for formation of the
self-regulated synchrotron cutoff :  
\begin{equation}
B_{100} \geq 0.8 \ R_{15}^{-2/3} \eta^{1/3}   \, ,
\end{equation}
where $R_{15}=R/10^{15} \, \rm cm$ is the source radius 
in units $10^{15} \, \rm cm$.

If  the synchrotron $\gamma$-rays are 
produced in the relativistically moving source,
the spectral cutoff is shifted towards the  TeV domain:  
\begin{equation}
\epsilon_0 \simeq 3  \ \delta_{10} \ \eta^{-1} \; \rm TeV \, .
\end{equation}
Thus, in the regime of acceleration 
with $\eta \leq 10$, the proton-synchrotron radiation  
emitted by a highly magnetized ($B \sim 100 \, \rm G$)
compact blob  with a typical size 
$R_{15}  \simeq 3.2   \Delta  t_{\rm 3h}  \delta_{10} \; \rm cm$  
and   Doppler factor $\delta_{\rm j}  \geq 10$,  results in  effective 
production  of TeV $\gamma$-rays. Remarkably, despite possible changes of some 
principal model parameters,  first of all the size $R$ and the magnetic 
field $B$ of the production region (e.g. caused by  
expansion or compression of the blob), we should expect a 
rather stable position
of the synchrotron cutoff,   provided that the parameter $\eta$ 
and  the Doppler factor $\delta_{\rm j}$
remain unchanged.
Meanwhile any change of $B$ and/or $R$ should result in 
strong ($\propto B^2 R^3$) variations  of the absolute 
flux of synchrotron  radiation. 
This intrinsic feature of the proton-synchrotron radiation could explain 
in a natural way the effect of weak correlation
between the spectral shape and the absolute flux 
of Markarian 501 revealed during the extraordinary 
outburst of the source in 1997  (Aharonian et al. 1999a,b). 

Dramatic changes of conditions in the $\gamma$-ray
production region should  lead, in fact, to the  transition
of the source from the regime dominated by synchrotron losses
to the regime dominated by the particle escape or by adiabatic losses.
For example, in the case of relativistically expanding blob
in which the reduction of the magnetic field is  faster than 
$B \propto R^{-2/3}$,  at some stage of the source evolution 
the condition represented  by Eq.(20)  could be violated,
and thus the spectral shape of th proton-synchrotron radiation 
would become  time-dependent. Indeed, in this regime the cutoff
energy in the spectrum of protons is determined 
by the balance between $t_{\rm acc}$ and 
$t_{\rm esc}$,  which gives
\begin{equation}
\epsilon_{0} \simeq 5.2 \ B_{100}^3  
R_{15}^2  \eta^{-2} \delta_{10} \,   \rm \, TeV \, .
\end{equation}

Note that Eq.~(20) keeps the synchrotron cutoff  energy,  
which is formed in the 
regime dominated  by particle escape,   
always less than the cutoff energy  formed in the 
synchrotron-loss-dominated  regime. For example,  
at  stages of source evolution when   
$R \leq 3 \times 10^{15} \ \rm cm$  
and  magnetic field $B \leq  10 \, \rm G$,
the $\gamma$-ray  spectrum is formed in the regime dominated 
by particle escape, thus
$\epsilon \leq  0.05 \eta^{-2} \ \delta_{10}  \rm TeV$, and correspondingly
at energies beyond 1 TeV the $\gamma$-ray spectrum  
becomes very steep.
In the {\em escape-loss-dominated} regime the cutoff energy,
and thus the shape of the entire synchrotron spectrum, strongly depend on
the magnetic field and the size of the source. 
In these stages of source evolution a significant  
spectral variability is expected
Depending on the relationship between the three appropriate
time-scales, $t_{\rm acc}$, $t_{\rm sy}$, and $t^{\ast}$,
we may predict,  similar  to the electron synchrotron radiation
in blazars (Kirk et al. 1998), quite different variability patterns such as
``soft lags'' (when $t_{\rm sy} \gg  t^{\ast} \gg  t_{\rm acc}$),
``hard lags'' (when $t_{\rm sy} \sim  t_{\rm acc} \sim  t^{\ast}$), etc.     

Finally, it should be noted  that the standard
self-regulated   synchrotron cutoff determined 
by  Eq. (19) corresponds to  the scenario 
when the proton acceleration and $\gamma$-ray production 
take place in the {\em same} localized region.
In the case of external injection of accelerated protons into 
the magnetized cloud, the energy of synchrotron 
radiation is determined  by the magnetic 
field in the $\gamma$-ray production region  
and the maximum energy of injected protons, 
therefore it could  exceed the 
`self-regulated synchrotron  cutoff.  
 
\section{Implications to Markarian 501 and Markarian 421}
The  HEGRA and Whipple observations of Markarian 501 during the 
extraordinary outburst in 1997  revealed  that  despite spectacular 
(up to factor  of 10 or more)   flux variations,  the shapes of the daily 
$\gamma$-ray spectra remained  unchanged\footnote
{In fact,  the CAT group has  found
a tendency  of the spectrum of Markarian~501 to become   
somewhat harder during the  flares (Djannati-Atai et al. 1999). However, 
the effect is  not very  strong and could not change   
the general conclusion about {\it essentially} stable spectral shape 
of TeV radiation in the high state.} 
throughout the entire state of high 
activity (Aharonian et al. 1999a,b; Quinn et al. 1999).  
To some extent this could be true also for Markarian 421 
(Aharonian et al. 1999c; Krennrich et al. 1999), 
although the uncertainties in the  
daily TeV spectra are too large for a firm conclusion.
  
\begin{figure}[htbp]
\begin{center}
\includegraphics[width=0.75\linewidth]{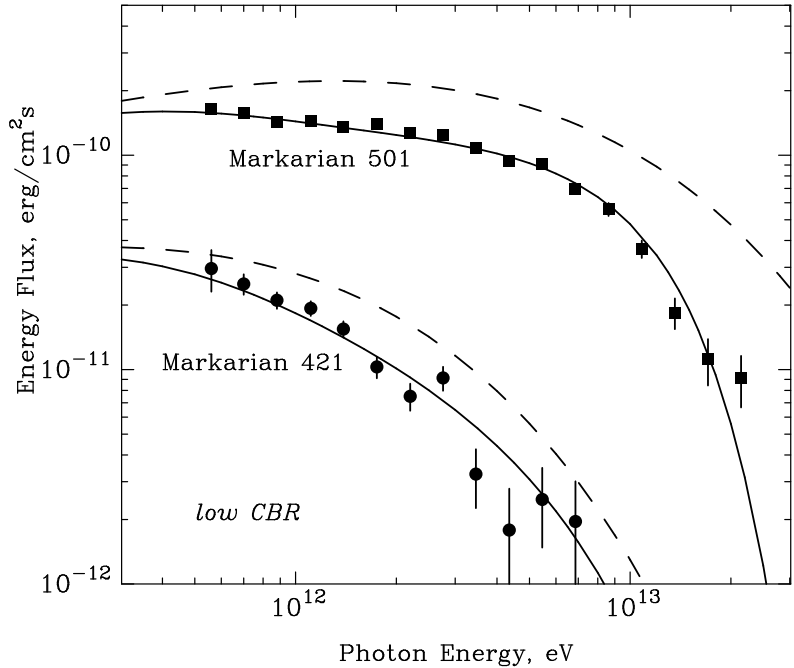}  
\caption{The proton-synchrotron radiation fits to 
the TeV spectra  of Markarian 501 and
Markarian 421. The TeV data of  Markarian 501 correspond to
the {\em high-state} spectrum of the source as measured by HEGRA 
in 1997  (Aharonian et al. 1999b).   
The TeV data  of Markarian 421 correspond to the  
average HEGRA spectrum based on the 
measurements in 1997-1998 (Aharonian et al. 1999c). 
The dashed and solid curves correspond to 
the spectra of proton-synchrotron radiation before and after 
correction for intergalactic extinction, respectively. A ``low-CBR''
model is assumed (solid curve in Fig.~6). 
The data of Markarian 501 are fitted assuming
$\alpha_{\rm p}=2$ and $\epsilon_0=1.3 \, \rm TeV$,  
and the data of Markarian 421 are fitted 
assuming $\alpha_{\rm p}=2$ and $\epsilon_0=0.26 \, \rm TeV$.} 
\end{center}
\end{figure}

Markarian 421 and Markarian 501  exhibit substantially 
different $\gamma$-ray spectra.
The time-averaged spectra of  
Markarian 501 in 1997  (Aharonian et al. 1999b) and Markarian 421 
(Aharonian et al. 1999c) as measured by the HEGRA system of telescopes
are shown in Fig.~5.  The differential flux of Markarian 501   
from 500 GeV to 24 TeV is described  
by a power-law with an exponential cutoff:
${\rm d} N/{\rm d} \epsilon \propto \epsilon^{-\Gamma} 
\, \exp{(-\epsilon/\epsilon_0)}$ with $\Gamma \simeq 2$ and
$\epsilon_0 \simeq 6.2 \, \rm TeV$ (Aharonian et al., 1999b). 
Remarkably, a quite similar spectrum was detected by the HEGRA
collaboration during a short, but strong outburst  of Markarian 501 
in June 1998 (Sambruna et al. 2000).  
In the TeV regime,  Markarian 421 has a  steeper spectrum. 
In the interval 
from 500 GeV to 7 TeV it could be approximated  by a pure power-law 
with a photon index $\Gamma \simeq 3$ (Aharonian et al. 1999c;
Piron et al. 2000),  or by  a power-law with 
$\Gamma=2.5 \pm 0.4$ and exponential cutoff at 
$\epsilon_0=2.8^{+2.0}_{-0.9} \, \rm TeV$  (Aharonian et al. 1999c).

\subsection{Extragalactic extinction of gamma rays}

The spectra of TeV radiation observed from distant
($d \geq 100 \, \rm Mpc$) extragalactic objects
suffer essential  deformation  during the passage
through the intergalactic medium   
caused by energy-dependent absorption  of 
primary $\gamma$-rays at interactions with  
the diffuse extragalactic background radiation  
(Gould \& Schreder 1966; Stecker et al. 1992, 
Vassiliev 2000).
For Markarian 421 and Markarian 501 at distances
150 Mpc and 170 Mpc, respectively, this effect could be
significant already at sub-TeV energies (Guy et al. 2000), 
and becomes especially  strong at energies above 10 TeV,  
where the optical depth $\tau_{\gamma \gamma}$  most likely 
significantly exceeds 1  (Coppi \& Aharonian  1999b). 
The lack of a relevant broad-band information
about the cosmic background radiation (CBR) from  
sub-micron to  sub-mm wavelengths 
introduces an ambiguity in the interpretation 
of the observed $\gamma$-ray spectra. 
Nevertheless,  the recent  reports about  
detection of CBR at near infrared 
($2.2 \, \mu \rm m$ and  $3.5 \, \mu \rm m$ - Dwek \& Arendt 1998;
Gorjian et al. 2000), 
mid infrared  ($15 \, \mu \rm m$ - Elbaz et al.  1999), and far 
infrared ($140 \, \mu \rm m$ and  $240 \, \mu \rm m$ -
Schlegel et al. 1998; Hauser et al. 1998) bands, 
allow significant reduction of the uncertainty 
in the intergalactic extinction.  Yet, these measurements {\em alone} 
are not sufficient for quantitative study of the 
intergalactic extinction of $\gamma$-rays, 
$\kappa(E)=\exp{[-\tau_{\gamma \gamma}(\epsilon)]}$,
in the observed energy range from  0.5 to 20 TeV.
Apparently, a suitable theoretical model of CBR is needed
which should reproduce properly the shape
of both ``stellar'' and ``dust'' components of 
the background radiation. Note that most of 
cosmological models give rather similar shapes of CBR
with two distinct bumps at 1-2 $\mu \, \rm m$ and 
100 -200 $\mu \, \rm m$ and a valley at 10-20  $\mu \, \rm m$
(see, e.g., Dwek et al. 1998; Primack et al. 1999).

\begin{figure}[htbp]
\begin{center}
\includegraphics[width=0.75\linewidth]{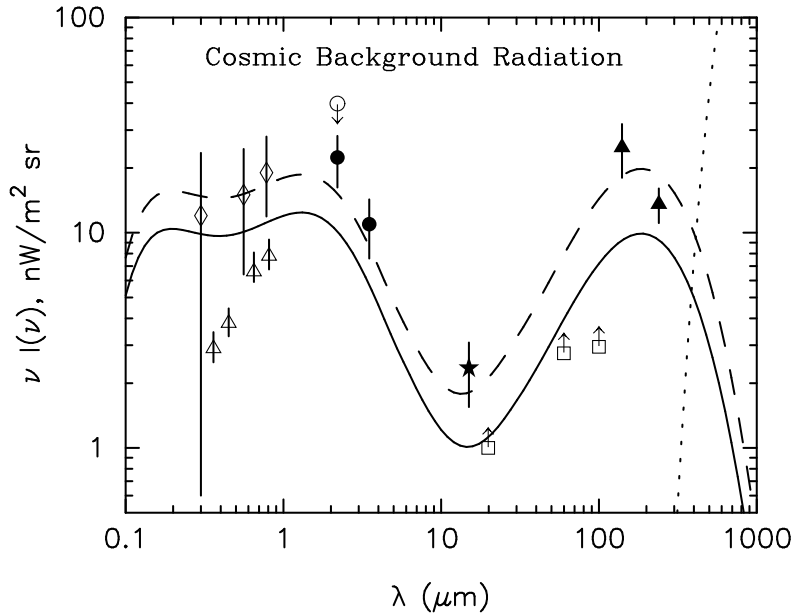}  
\caption{The Cosmic Background Radiation (CBR). 
The solid curve corresponds to the 
``low-CBR'', and the dashed curve  corresponds to the ``high-CBR''  model
predictions (see the text).  The dotted curve shows  the flux
of the 2.7 K microwave background radiation.  
The tentative detections of CBR at $2.2 \mu \rm m$ and $3.5 \mu \rm m$
(filled circles)  are from Gorjian et al. (2000), 
and at $15 \mu \rm m$ (the star) 
from  Elbaz et al. (1999); the fluxes at $140 \mu \rm m$ and $240 \mu \rm m$
(filled triangles) are from  Hauser et al. (1998).
The other upper/lower CBR flux limits (open symbols) are taken from the recent
compilation of CBR by Dwek et al. (1998).} 
\end{center}
\end{figure}

Below I will use the so-called LCDM model 
of Primack et al. (1999), but the absolute fluxes 
of both ``stellar'' and ``dust''components of radiation
will be allowed to vary 
within a factor of 2 or so. The prediction for the CBR flux by the LCDM
model (hereafter ``low-CBR'') and  re-scaled LCDM model, assuming  
50 per cent  higher flux of the  ``stellar'' component,  and twice  
higher flux of  the ``dust'' component  (hereafter ``high-CBR''),  
are presented  in Fig.~6. For comparison,  the recently reported  CBR flux 
measurements or upper/lower  limits from UV to FIR are also shown. 
Despite large statistical and 
systematic uncertainties, the actual  
spectrum of CBR  most probably does not  deviate significantly 
from the two simplified  model predictions shown in Fig.~6. 
 
\subsection{Fitting the TeV spectra of Markarian 501 and Markarian 421}
The  spectrum of the proton-synchrotron radiation 
corrected for intergalactic extinction
can satisfactorily  fit, for a reasonable combinations of a limited number 
of model parameters,  the observed TeV fluxes  of both  
Markarian 501 and Markarian 421.  
This is demonstrated in Fig.~5, where the average TeV  
fluxes of both objects  are shown together with the 
model predictions.  The theoretical spectra, after corrections for the 
intergalactic extinction assuming  the ``low CBR'', 
are normalized to the  measured fluxes at 
1 TeV.  This determines,  for the given magnetic field,  
the required total energy in accelerated protons.
Assuming that the accelerated protons 
have an energy distribution represented by 
Eq.~(7),  the observed spectra of TeV emission  
can be fitted  by the proton-synchrotron radiation for
the following combinations of model parameters:  
$\alpha_{\rm p}=2$ and $\epsilon_0=1.3 \, \rm TeV$ for Markarian 501, 
and $\alpha_{\rm p}=2$ and $\epsilon_0=0.26 \, \rm TeV$  
for Markarian 421. Assuming now that the $\gamma$-rays 
are produced in the  {\em synchrotron-loss-dominated} regime, 
from Eq.~(21) one can easily estimate the 
ratio $\rho=\delta_{10}/\eta$  which is the most relevant
parameter for determination of the position of the self-regulated 
synchrotron cutoff -- $\rho=0.43$ and $0.09$
for Markarian 501 and Markarian 421, respectively.  
 
The predicted for  Markarian 501 proton-synchrotron radiation flux 
passes below the HEGRA point at 24 TeV (see Fig.~5). 
This discrepancy could be removed, in principle, by assuming 
a different (steeper)  shape of the spectrum of the ``dust'' component 
compared with  predictions 
of current  CBR models (Coppi \& Aharonian 1999b), and/or 
assuming a flatter spectrum of protons in the region of the cutoff
$E_0$ compared to  the spectrum given by Eq.~(7). 
However, because of the inadequate statistical significance 
of the $\gamma$-ray  data above 16 TeV 
(Aharonian et al. 1999b),  the last point  on Fig.~5 
should be considered a flux upper limit rather 
than a positive detection.

\begin{figure}[htbp]
\begin{center}
\includegraphics[width=0.75\linewidth]{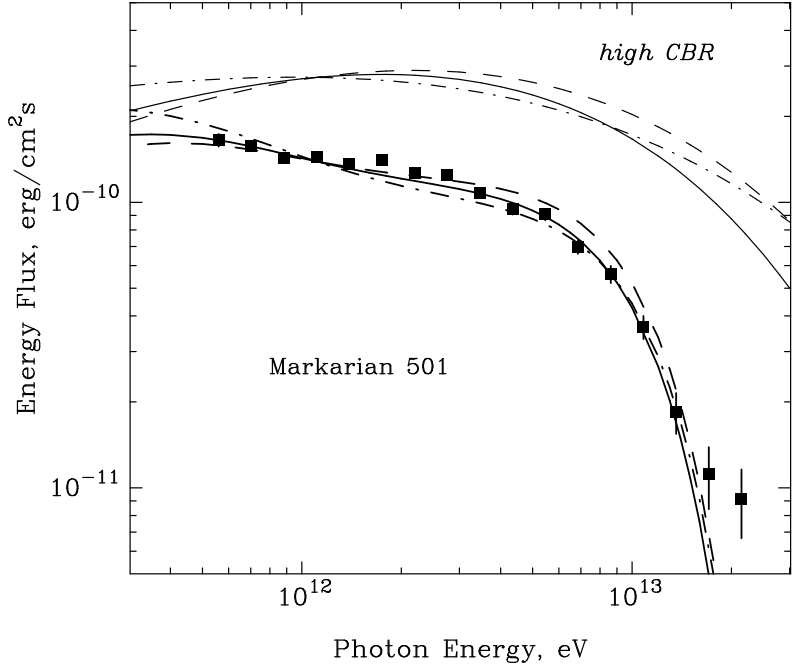}  
\caption{The SED of Markarian 501.
The thin  and heavy  curves 
correspond to the spectra of proton-synchrotron radiation before and after 
correction for the intergalactic extinction, respectively. The  ``high-CBR''
model is assumed (dashed curve on Fig.~6). 
The data of Markarian 501 are fitted by
parameters $\alpha_{\rm p}=2$ and $\epsilon_0=1.8 \, \rm TeV$ (solid curve),
$\alpha_{\rm p}=1.5$ and $\epsilon_0=1.2 \, \rm TeV$ (dashed curve), and
$\alpha_{\rm p}=2.4$ and $\epsilon_0=3.5 \, \rm TeV$ (dot-dashed curve).
} 
\end{center}
\end{figure}
					
If one treats  the TeV observations independently  of data 
obtained  in  other energy domains, the  parameters chosen 
to fit the TeV spectra cannot be unique in a sense that
some  other combinations of $\alpha_{\rm p}$ and  $\epsilon_0$ 
could  still provide reasonable  spectral fits.
This is demonstrated in Fig.~7  where  
the measured fluxes of Markarian 501  are shown together 
with  the predicted spectra of the proton-synchrotron radiation 
assuming ``high-CBR'' model for the background radiation. 
It is seen that the detected fluxes could be equally well fitted by 
synchrotron radiation for 3 different combinations: 
(i) $\alpha_{\rm p}=2$ and 
$\epsilon_0=1.8 \, \rm TeV$ (or $\rho \simeq 0.6$,  
if the energy losses of protons are dominated by the 
synchrotron radiation), 
(ii) $\alpha_{\rm p}=1.5$ and 
$\epsilon_0=1.2 \, \rm TeV$ ($\rho \simeq 0.4$),
and  $\alpha_{\rm p}=2.4$ and $\epsilon_0=3.6 \, \rm TeV$  
($\rho \simeq 1.2$).
The impact  of the  uncertainty in the CBR flux 
can be estimated  by comparing the solid curves 
in Fig.~5 and Fig.~7. While in both cases 
the power-law index of the proton spectrum is assumed
$\alpha_{\rm p}=2$, the  ``high-CBR''
requires somewhat ($\approx 40$ per cent) higher synchrotron 
cutoff compared with the case of ``low-CBR''. Also,
for the ``high-CBR'' the discrepancy of the 
theoretical and measured fluxes above
16 TeV becomes more apparent.   
 
An important  constraint on the possible  
``$\alpha_{\rm p}$/$\epsilon_0$''  
combination  could be provided  by $\gamma$-ray data obtained 
at low, MeV/GeV energies. In particular, the combination of 
parameters used  for the fit of the spectrum of   Markarian 421 
in  Fig.~5,  predicts  an energy flux at 100 MeV of about 
$2 \times 10^{-11} \, \rm erg/s$, i.e. factor of 3 less than 
the EGRET flux averaged over the 1992-1996 
period of observations (Sreekumar et al. 1996).  Therefore,  assuming that 
the observed MeV/GeV fluxes are associated with the 
proton-synchrotron radiation  
as well, the best fit of the data from 100 MeV to 10 TeV 
could be obtained adopting  a steeper proton spectrum 
with $\alpha_{\rm p}=2.4$ (see Fig.~8).   
Although the data obtained at TeV and MeV/GeV energies 
correspond to different epochs,
the lack of evidence of strong spectral variability   
in both energy domains (Aharonian et al. 1999c; 
Sreekumar et al. 1996) makes  such a fit quite meaningful.

\begin{figure}[htbp]
\begin{center}
\includegraphics[width=0.65\linewidth]{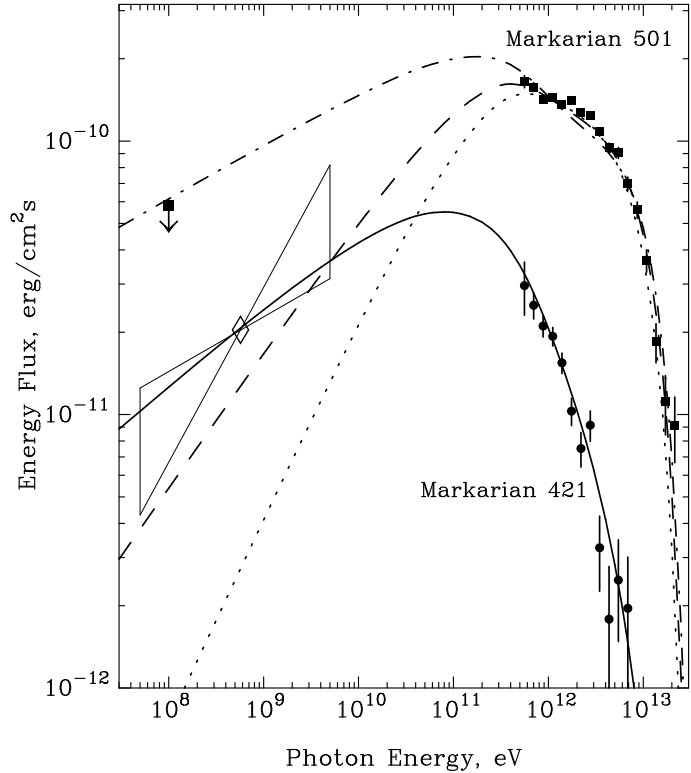}  
\caption{The proton-synchrotron radiation of Markarian 501 and
Markarian 421 from MeV/GeV to TeV energies. All spectra
are  normalized to the observed fluxes at 1 TeV after  
correction  for the  intergalactic extinction assuming the 
``low-CBR''  model.  The data of Markarian 421 are fitted
by $\alpha_{\rm p}=2.4$ and $\epsilon_0=1 \, \rm TeV$. 
The data of Markarian 501 are  fitted by three 
different combinations of $\alpha_{\rm p}$ and 
$\epsilon_0$: 
$\alpha_{\rm p}=2$ and $\epsilon_0=1.3 \, \rm TeV$  (dashed curve); 
$\alpha_{\rm p}=2.6$ and $\epsilon_0=4 \, \rm TeV$ 
(dot-dashed curve); $\alpha_{\rm p}=1.5$ and $\epsilon_0=0.75 \, \rm TeV$ 
(dotted curve).  The TeV data of  Markarian 501 and 
Markarian 421 are the same as in Fig~2. 
The filled square at 100 MeV  
corresponds to the flux upper limit set by EGRET during
the period  April 9-15 (1997) when  Markarian 501 
was in a very high state (Catanese et al. 1997).    
The zone of the low-energy $\gamma$-ray fluxes of Markarian 421  
shown from 50 MeV to 5 GeV corresponds to the spectrum of the source 
averaged over the Phase-1 period of the 
EGERET observations (Sreekumar et al. 1996).} 
\end{center}
\end{figure}

The observations of Markarian 501 by EGRET during the
extraordinary  active state in April 1997 
resulted  only in the   
flux upper limit at 100 MeV (Catanese et al. 1997).
This upper limit, combined with the TeV data,   
constrains the power-law index of the proton spectrum,
$ \alpha_{\rm p} \leq 2.6$.  For $ \alpha_{\rm p} = 2.6$  the 
position of the synchrotron cutoff should appear at  
high energies, $\epsilon_0=4 \, \rm TeV$, 
in order to match the  TeV spectrum  (see Fig.~8).   
For the very flat proton spectrum   
with $\alpha_{\rm p}=1.5$,  the synchrotron cutoff occurs 
below 1 TeV, $\epsilon_0 \simeq 0.75 \, \rm TeV$,   
otherwise the calculated  spectrum  becomes harder than 
the observed TeV spectrum. And finally, for the ``standard''
proton spectrum with $\alpha_{\rm p}=2$,  the best fit value for 
the synchrotron cutoff is  provided by  
$\epsilon_0 \simeq 1.3 \, \rm TeV$,   or $\rho \simeq 0.43$
if the TeV   spectrum of  Markarian 501  is 
formed  in the synchrotron-loss-dominated regime. 
Because  the Doppler factor of the small-scale jet in 
Markarian 501  is believed to be close to 10,  a conclusion
could be drawn  that  during the entire 1997 
outburst the acceleration of particles 
in Markarian 501 took place in the regime  of  maximum 
acceleration rate ($\eta \sim  2$). If so, this explains in a rather natural way
the essentially time-independent shape of the 
TeV spectrum observed during the strong flares 
in 1997 (Aharonian et al. 1999a,b).

Meanwhile, a significant  drop in   the acceleration rate 
(and/or the  strength of the magnetic field)   should lead to dramatic 
changes of the spectrum of $\gamma$-rays.  
This effect perhaps can explain the recent 
report of detection by EGRET of a very flat $\gamma$-ray flux  at energies
from 50 MeV to 5 GeV during the period  of  relatively low state of 
Markarian 501 in  March 1996.  A  rather surprising aspect of this report  
is that  the energy flux of  $\gamma$-rays at several  GeV  was  
significantly larger  than the X-ray and TeV fluxes 
measured approximately at the same  time by ASCA and Whipple 
instruments (Kataoka et al. 1999). 
In the  framework of the  ``proton-synchrotron radiation''  hypothesis,
the  EGRET and Whipple fluxes  could be explained 
with  model parameters $\alpha_{\rm p}=2$ and  $\epsilon_0=0.02 \, \rm TeV$, or,
if the proton acceleration takes place in  the regime dominated 
by synchrotron losses,  
$\rho \simeq 0.007$ (Fig.~9). The latter implies dramatic
reduction of the acceleration efficiency 
($\eta \simeq 140 \delta_{10}$) compared with the high state, 
provided that  the jet's Doppler factor remains 
more or less constant.
It is necessary to notice that this interpretation
which adopts very small $\rho$ parameter, but 
tacitly assumes  that the proton acceleration takes place in the 
{\em synchrotron-loss-dominated} regime,  requires, 
as it follows from Eq.~(20),  an extremely large magnetic field,  
$B \geq  400 R_{15}^{-2/3} \delta_{10}^{1/3} \, \rm G$. 
A more   realistic scenario seems the transition of the 
source from the  {\em synchrotron-loss-dominated} regime to the 
{\em particle escape-loss-dominated} regime, for example,  due to a 
possible reduction of   the magnetic field  caused by 
the expansion of the blob.
The significant drop of the magnetic field,
e.g. from  $B \sim 100 \, \rm G$ to  $B \sim 10 \, \rm G$
would result in an earlier synchrotron  cut-off.   

\begin{figure}[htbp]
\begin{center}
\includegraphics[width=0.60\linewidth]{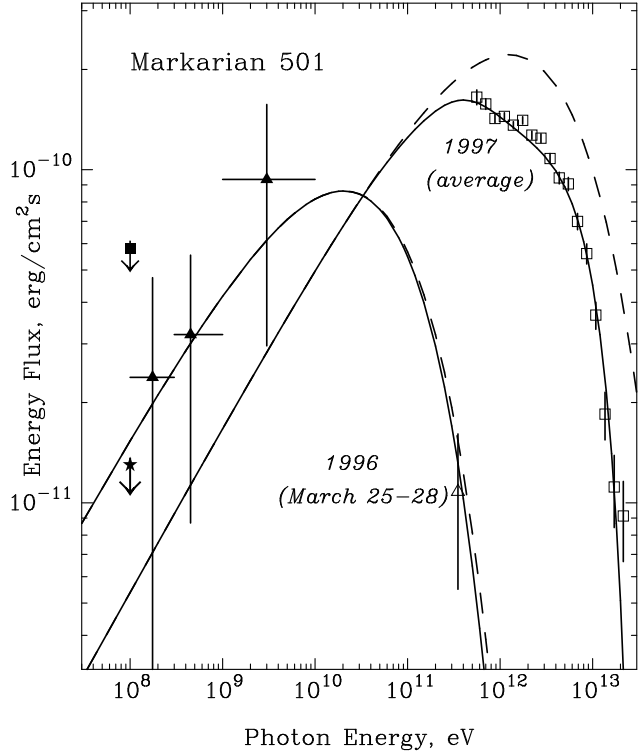}  
\caption{The  proton-synchrotron radiation of Markarian 501 
in a high and low states. The dashed and solid curves correspond to 
the spectra of proton-synchrotron radiation before and after 
correction for intergalactic extinction, respectively. The ``low-CBR''
model is assumed. 
The data of Markarian 501 in the high state are the same as in 
Fig.~2. The broad-band $\gamma$-ray data in the low state 
are obtained  by the Whipple (open triangle) and EGRET (filled triangles) 
groups  during the multiwavelength campaign in 
March 1996  (Kataoka et al. 1999).  An
``archival'' upper limit on the 100 MeV flux based on the long-term 
observations  of the source during the phase I period of  EGRET is also 
shown (star). The high state of the source is fitted by 
$\alpha_{\rm p}=2$ and $\epsilon_0=1.3 \, \rm TeV$. 
The low state of the source is fitted by 
$\alpha_{\rm p}=2$ and $\epsilon_0=0.02 \, \rm TeV$.}
\end{center}
\end{figure}

The signature of such a low state is the shift of the position
of the peak of SED  down to   100 GeV or below, with
a hard spectrum of GeV $\gamma$-rays, and very 
steep spectrum at TeV energies.  While such a 
hard spectrum of GeV radiation is indeed observed, 
the small TeV photon statistics  
do not provide an adequate  information 
about the spectrum at highest energies. Contrary to the 
{\em synchrotron-loss-dominated regime},  the spectrum of 
$\gamma$-rays   formed in the {\em escape-loss-dominated regime}
depends on parameters characterizing the production
region, in particular the source size and the magnetic field.
Therefore, in this stage of evolution,  the source should show 
significant variations  of  the spectral shape of TeV 
$\gamma$-ray emission.

The TeV spectrum of Markarian~421 is significantly steeper 
than the spectrum of Markarian~501. Another
distinct feature of Markarian 421 is the TeV  
flux variability on extremely short, $\Delta t \sim 15$  
minute timescales  (Gaidos et al. 1996). 
Assuming that the radiation 
is formed in the synchrotron-loss-dominated 
regime, the steep $\gamma$-ray spectrum  of Markarian 421 
with  $\epsilon_0 \simeq 0.26 \, \rm TeV$ (see Fig.~5)  
requires relatively slow acceleration rate, 
$\eta \simeq 12 \delta_{10}$. On the other hand, in this regime
the radiative cooling time of protons responsible for 
production of $\gamma$-rays 
in the cutoff region, $t_{\rm sy} \simeq 2.4 \times 
10^4 \ B_{100}^{-3/2} \eta^{1/2}$ s,
could match the observed variability  
(in the frame of the jet)  
$9 \times 10^3 \delta_{10} \, \rm s$,  if
$B_{100} \geq 2 \eta^{1/3} \delta_{10}^{-2/3}$.
Therefore, $B_{100} \geq 4.5 \ \delta_{10}^{-1/3}$.  
Thus,  if the TeV radiation of  Markarian 421 is 
formed in the  {\em synchrotron-loss-dominated regime}, 
the magnetic field should  significantly exceed 100 G, 
which seems a rather extreme assumption. 

In a relatively moderate (and perhaps more realistic) 
magnetic field of about 100 G or less, both the {\em observed 
rapid variability} and the {\em steep TeV spectrum}  of Markarian 421 
could be interpreted, assuming that the radiation is formed 
in the {\em escape-loss-dominated regime}. Indeed, in the compact 
$\gamma$-ray production region with 
$R \leq c \ \Delta t \ \delta_{\rm j} \simeq 
2.7 \times 10^{14} \delta_{10} \, \rm cm$, 
the position of the synchrotron cutoff  
is estimated 
$\epsilon_0 \simeq 0.38 \ B_{100}^3 \delta_{10}^3 \eta^{-2} \, \rm TeV$
(see Eq.~22). For a reasonable combination 
of parameters  $B_{100} \sim 1$, $\delta_{10} \sim 1$ and 
$\eta \sim 1$, this estimate agrees quite well with the
allowed range of the synchrotron cutoff between 0.25 and 1 TeV as 
derived from the TeV spectrum of Markarian 421 (Fig.~5 and Fig.~8). 

\subsection{Synchrotron radiation from the 
proton distribution with sharp pile-up}

The spectra of the proton-synchrotron radiation discussed above 
are calculated under the assumption of 
a  ``standard '' (power-law with exponential cutoff) 
proton spectrum represented by Eq.~(7). Actually, the spectra of 
accelerated particles at some circumstances 
could have  more complicated and exotic forms, e.g. 
they could contain  pile-ups
preceding the spectral cutoffs.  An example of a 
SED of synchrotron radiation corresponding
to the proton spectrum with sharp pile-up and abrupt cutoff
is shown in Fig.~10. The shape of this spectrum (dashed curve) 
essentially differs from the
measured spectrum of Markarian 501. Moreover,  
for any reasonable spectral distribution
of the CBR, the deformation of the initial $\gamma$-ray
spectrum caused by the intergalactic extinction, hardly
could bring it closer to the observed one. The desirable shape, 
however, could be achieved,  
if we assume that the deformation of the primary $\gamma$-ray
spectrum takes place in two steps, e.g. presuming  that 
the primary radiation suffers significant absorption {\em twice} - 
firstly in  the vicinity of the source, and afterwards  - in the
intergalactic medium.  The synchrotron radiation of the 
jet itself seems  to be a natural photon field to serve as 
an internal absorber for the TeV radiation. Actually, 
such absorption is even unavoidable if the Doppler factor of the jet
$\delta_{\rm j} \leq 10$. However, the 
pronounced peak in the {\it production} spectrum of 
radiation shown in Fig.~10 (dashed curve), 
requires rather selective absorption with strongest  impact 
in the region of the $\gamma$-ray peak, which  cannot be  provided by
the broad and flat spectrum of low-frequency 
synchrotron radiation of the jet. Because 
$\gamma \gamma \rightarrow e^{+}e^{-}$ cross-section 
peaks near the threshold, there is approximately a one-to-one
correspondence between the energies of the 
absorbed $\gamma$-ray $\epsilon^\ast$ and the field photon $h \nu$:
$\epsilon^\ast \simeq 1  \ (h \nu/1 \, \rm eV)^{-1} \, \rm TeV$.
Therefore, for a selective absorption of $\gamma$-rays 
from the region of the maximum of the production  spectrum,
the target photon field should have a quite  narrow distribution, 
$\Delta\nu/\nu \sim 1$. 
%
\begin{figure}[htbp]
\begin{center}
\includegraphics[width=0.75\linewidth]{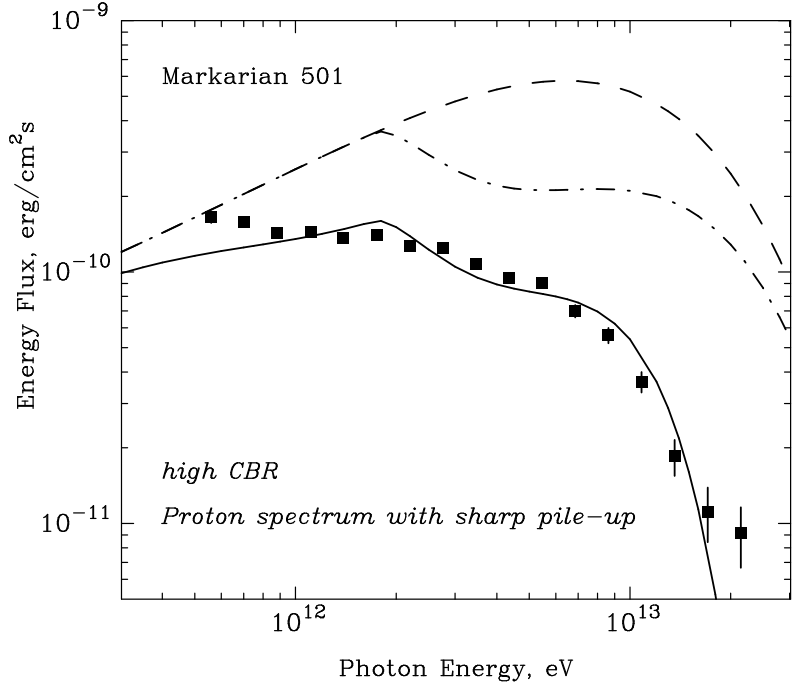}  
\caption{The proton-synchrotron radiation of 
Markarian 501 calculated for the proton spectrum with
a sharp pile-up (curve 3 in Fig.~1).
Dashed curve - the production spectrum; dot-dashed curve - 
$\gamma$-ray spectrum after internal absorption; 
solid curve - $\gamma$-ray spectrum after internal and 
intergalactic absorptions. The following parameters 
are assumed: 
$\epsilon_0=7.8 \, \rm TeV$ (or $\rho=2.6$), the mean energy of
target photons responsible for internal absorption 
$\overline{h \nu}=0.15 \, \rm eV$, optical depth 
$\tau_{\gamma \gamma}=1$ at 
$\epsilon =7 \, \rm TeV$.  For the extragalactic extinction 
of $\gamma$-rays the ``high-CBR'' model is assumed.
}
\end{center}
\end{figure}	
The dot-dashed curve in Fig.~10 demonstrates the effect of 
such absorption in a monochromatic photon field with 
energy $h \nu=0.15 \, \rm eV$, and assuming that  
$\tau_{\gamma \gamma}=1$ at $\epsilon \simeq 7 \, \rm TeV$. 
A more realistic, e.g.  thermal (Plankian) distribution of the target photons
with temperature $kT \simeq 1/3 \ h \nu \simeq 0.05 \, \rm 
eV$ gives a similar result,  with a bit smoother
spectrum of absorbed $\gamma$-rays at 2 TeV. The optical depth 
$\tau_{\gamma \gamma} \sim 1$ caused by  mid infrared  
($\lambda \sim 10 \mu \rm m$) photons,  could be provided 
by different kind of  sources  in the vicinity of the central black hole 
(Celotti et al. 1998).

The TeV $\gamma$-rays, after suffering additional absorption 
in the intergalactic photon fields,  will eventually  arrive  
with ``double-deformed'' spectrum shown in Fig.~10 (solid curve). 
The calculated synchrotron spectrum, after corrections for 
internal  and intergalactic absorption and normalized to the 
observed flux at 1 TeV,  quite satisfactorily 
fits the SED of Markarian 501 in the high state,
taking into account significant  (up to factor of 2) systematic
uncertainties of the HEGRA flux measurements 
at energies below 1 TeV and above 15 TeV (Aharonian et al., 1999b). 

\section{Discussion}

Although there is  little doubt 
that the bulk of the  highly variable X-ray emission of the 
BL Lac objects has  synchrotron origin and is produced  in the relativistic 
jets pointed to the observer (Urry \& Padovani 1995), the TeV radiation of 
Markarian 421 and Markarian 501 is the only model-independent 
and unambiguous indicator of acceleration of ultra-relativistic 
particles in these small-scale  (sub-pc) jets.
In the currently popular synchrotron-self-Compton (SSC) models 
of BL Lac objects both the X-ray and $\gamma$-ray components are 
attributed to the radiation of directly accelerated TeV electrons.
For a certain, physically well  justified combination of 
parameters (the blob size, magnetic field, Doppler factor, etc.), 
these models (see, e.g., Inoue \& Takahara 1996, 
Mastichiadis \& Kirk 1997, Pian et al. 1998, 
Bednarek \& Protheroe 1999, Coppi \& Aharonian 1999a,
Wagner et al. 1999; Kataoka et al. 1999,
Maraschi et al. 1999, Takahashi et al. 1999,  
Krawczynski et al. 2000)
give quite  satisfactory explanation of the observed 
spectral and temporal characteristics of non-thermal radiation 
over more than ten decades  of frequencies  from 
$10^{16} \, \rm Hz$ to  $10^{27} \, \rm Hz$.
In the SSC  models  the strength of the
magnetic field in  the $\gamma$-ray emitting regions  typically 
is less than $1 \, \rm G$,  the most likely value being 
close to $0.1 \, \rm G$, which  
directly follows from the comparable energy fluxes released 
in  X-rays and $\gamma$-rays  during the TeV  flares of  
Markarian 421 and Markarian 501 (e.g. Tavecchio et al. 1998).  
At such a low magnetic field  the maximum energy of accelerated  
protons cannot exceed 
$E_{\rm p, max} \sim 10^{18}  \Delta t_{\rm 3h}  \delta_{10} \ (B/1 \, \rm G) 
\, \eta^{-1} \rm eV$, and therefore  the contribution of protons
in $\gamma$-ray production through both the photo-meson and 
the synchrotron channels  is  negligible;   
the production of TeV  $\gamma$-rays in the jet is strongly 
dominated by the inverse Compton scattering of electrons directly
accelerated up to energies $\sim 10 \, \rm TeV$.  

The increase of the 
magnetic field leads to reduction ($\propto B^{-2}$) of  
the flux of 
the inverse  Compton $\gamma$-rays;  for $B \geq 1 \, \rm G$ 
the bulk of energy of accelerated electrons is channeled into the 
synchrotron radiation.  Also, the increase of the magnetic field 
shifts ($\propto B$)  the synchrotron peak to higher energies. 
But  for any reasonable Doppler factor of the jet, 
the observed TeV fluxes of Markarian 421 and Markarian 501 
cannot be explained by  the electron synchrotron radiation because of 
the self-regulated  synchrotron cutoff  
at $\epsilon_{\rm max} \simeq 1.6 \ \delta_{10} \, \rm  GeV$,  
which inevitably appears   if the electron acceleration  
and $\gamma$-ray production take place in the same region of the jet.
Formally, we may avoid this limit  assuming that the regions 
of the electron acceleration and the $\gamma$-ray production 
are separated, i.e. the electrons are accelerated 
up to energies $\sim 10^3 \, \rm TeV$ in a region 
with rather small magnetic field, 
$B \leq 0.01 \, \rm G$, but release all their energy in a  form of  
GeV/TeV synchrotron radiation later, after entering the region(s) 
of strongly compressed/amplified  magnetic field, 
$B \geq 10 \, \rm G$. This scenario seems, however,  rather artificial.
 
Large magnetic fields, $B  \sim 100 \, \rm G$ or so, 
coupled with effective acceleration of protons at the 
maximum rate, may create very favorable conditions for
TeV $\gamma$-ray production.  Indeed,  at such conditions 
the synchrotron cooling of protons not only well 
dominates over other radiative and non-radiative losses (see Fig.~4), 
but also provides good fits to  the observed spectra of TeV radiation of 
Markarian 421 and Markarian 501. 
Remarkably, if the proton acceleration
takes place in the {\em synchrotron-loss-dominated} regime,  the spectral shape
of radiation depends only on the power-law index of accelerated protons,
$\alpha_{\rm p}$ and the parameter $\rho=\delta_{10}/\eta$, but not on the 
magnetic field and  the size of the $\gamma$-ray
emitting blobs, which  generally endure significant time-evolution.
Meanwhile, any change in the size and/or magnetic field 
of the evolving  blobs should lead to significant variation in the absolute 
flux of $\gamma$-rays. This effect  could give a natural explanation for
one of the remarkable features of strong flares of Markarian 501
(and, perhaps, also Markarian 421)  - 
the essentially stable spectral shape of
TeV radiation despite strong variation of the absolute flux observed 
on time scales  less than 1 day.   During the strong flares 
of Markarian 501 with a flat spectrum around 1 TeV,   
the parameter $\rho=\delta_{10}/\eta \sim 1$. 
The exact value of $\rho$ depends on the level of the
diffuse extragalactic background and  on the power-law index 
of accelerated protons. The uncertainty in this parameter,  despite significant  
uncertainties in the intergalactic extinction of TeV radiation, 
as well as the lack of adequate information about the $\gamma$-ray fluxes  
at MeV/GeV energies,  does not exceed a factor of three or so.  
Since the Doppler factor  in Markarian 501 is believed to be within 
10 to 30, the acceleration rate should be 
very high, $\eta=\rho^{-1} \  \delta_{10} \sim 1-3$. 
The {\em unusually high} GeV flux observed from Markarian 501 
during the multiwavelength campaign in   March 1996,  when 
the source was in a {\em very low} TeV state, could be  
explained  by a transition of the source from  
the {\em synchrotron-loss-dominated} regime 
to the  {\em escape-loss-dominated} regime,   caused, for example, by 
dramatic drop of the magnetic field.  In this regime 
we should expect  steeper TeV $\gamma$-ray spectra with a  
slope depending on the magnetic field and the size of the source.
Therefore, an important test of this hypothesis could be detection of 
spectral variability of $\gamma$-radiation in a low state of the source.

The significantly steeper 
TeV spectrum of Markarian 421 
requires early synchrotron cutoff,  
$\epsilon_0 \sim 0.25 - 1 \, \rm TeV$.
This could be interpreted as a result of 
formation of the synchrotron spectrum   
in the regime dominated by the
escape of protons from the acceleration region. 
For a magnetic field of about 100~G this would require  
a compact source with $R < 10^{15} \, \rm cm$.
Interestingly, such a small linear size of the 
$\gamma$-ray production region is supported independently
by the observed dramatic variations of the TeV flux of 
Markarian 421 on  timescales $\sim 15$ minutes. 

Within the model of proton synchrotron radiation, we 
may expect also synchrotron radiation  
produced by {\em directly accelerated} electrons - the 
counterparts of EHE protons. If the particle acceleration 
takes place in the synchrotron-loss-dominated regime,
the  self-regulated synchrotron cutoff  
of this component depends only on the parameter $\rho=\delta_{10}/\eta$,
namely $\epsilon_0 \simeq  1.6  \rho \, \rm GeV$, thus it  
correlates with the position of 
the  cutoff  in the proton synchrotron spectrum 
at $ \epsilon_0 \simeq 3   \rho \, \rm TeV$.
The ratio of the energy fluxes of 
these two components is  determined simply by the  ratio of  
non-thermal energy  channeled into the  accelerated protons 
and  electrons, $\dot{W_{\rm p}}/\dot{W_{\rm e}}$. In a strong 
magnetic field of about 100 G, the synchrotron cooling time of
electrons is shorter, almost  at all relativistic energies, 
than the  typical dynamical (e.g. light-crossing) times. 
This results in a well-established steady-state spectrum of electrons 
${\rm d}N/{\rm d}E \propto E^{-(\alpha_0+1)}$, 
provided that the power-law index of acceleration spectrum 
$\alpha_0 \geq 1$. Consequently, a pure power law
synchrotron spectrum  with a photon index  $(\alpha_0+2)/2$ 
would be formed. In particular, for  $\alpha_0=2$, we should expect a flat 
synchrotron SED ($\nu S_\nu=const$) from  
the optical/UV wavelengths to 
MeV/GeV  $\gamma$-rays. 
The broad-band spectra of both Markarian 421 and Markarian 501 do 
not agree with  such a pure power-law behavior; in fact, the SED
of both objects show pronounced synchrotron X-ray peaks.    
Therefore this (theoretically possible) population  of 
directly accelerated electrons - 
counterparts of the EHE protons - cannot be  responsible 
for the  bulk of X-ray emission.

The latter could be referred to electrons, 
produced, most probably,  in a different way,
and/or in other region(s)  of the jet. 
The  X-ray light curves of both  Markarian~421 
(e.g. Takahashi et al. 1999) and Markarian~501 
(e.g. Sambruna et al. 2000) show  so-called ``soft'' 
and ``hard'' lags. A possible interpretation of this effect
in terms of competing acceleration, radiative cooling, 
and escape timescales  of synchrotron-emitting electrons 
(Takahashi et al. 1996; Kirk et al. 1998), would   
require  that all these timescales 
are comparable with the light crossing time, 
$t=R/c \sim 10^{4} - 10^{5}$ s. 
The cooling time of an electron responsible for 
a synchrotron photon  of energy $\epsilon$ is 
$t^{\rm (e)}_{\rm sy} \simeq 
1.5  \times 10^3 (B/1 \, \rm G)^{-3/2} \ (\epsilon/1 \, \rm keV)^{1/2} \, \rm s$. 
Therefore in the X-ray production region the magnetic  field cannot, 
independent of specific model assumptions, significantly 
exceed $0.1 \, \rm G$.  Thus,  the hypothesis of 
proton-synchrotron origin of TeV radiation  
implies  that the production regions of 
TeV $\gamma$-rays  ($B \sim 100 \, \rm G$) and 
and  synchrotron X-rays ($B \sim 0.1 \, \rm G$) 
are  essentially different. Although in any 
reasonable scenario we may expect a non-negligible correlation
between the  X-ray and  TeV 
$\gamma$-ray fluxes, the spatial separation of the X-ray and 
TeV $\gamma$-ray production regions does not allow 
definite  predictions for such a correlation.    

In the proton-synchrotron model of TeV radiation of BL Lac
objects, two more components of X-radiation are expected. 
Firstly, in the field of about 100~G  the accelerated  
protons  of energy  $E \sim 10^{15} \, \rm eV$  {\em themselves} 
produce  synchrotron X-rays. However, the  contribution of this
component to the observed X-ray flux is  negligible. 
A much more prolific channel for X-ray production
connected (indirectly)  with the EHE protons,  can be provided
by electrons of non-acceleration origin, namely by 
secondary electrons produced at interactions of the primary 
TeV $\gamma$-rays with the ambient low-frequency radiation.  
Actually,  for a spectrum of 
EHE protons containing sharp spectral pile-up, 
we {\em must} assume an essential 
($\tau_{\gamma \gamma} \sim 1$) internal absorption 
of $\gamma$-rays in order to
match the observed spectrum of Markarian 501 (see Fig.~10).
The appearance of secondary  electrons in the jet results 
in production of a hard synchrotron X-ray component. 
Apparently, for an optical depth  $\tau_{\gamma \gamma} \sim 1$,  
the luminosity of this component would be comparable to 
to the luminosity of  their ``grandparents'' - TeV $\gamma$-rays.   
The photo-produced electrons have a rather specific, 
significantly different from the  directly accelerated particles, 
shape. For example, the spectrum of electrons produced at interactions 
of  high energy $\gamma$-rays with a photon index $\Gamma$
and  field photons with a narrow (e.g. Planckian) 
spectral distribution  with a characteristic energy $\overline{h \nu}$,
has the following characteristic form: 
starting from the
minimum (allowed by kinematics) energy at 
$E_{\ast}=m_{\rm e}^2 c^4/4 \overline{h \nu}$, the electron spectrum sharply 
rises reaching the maximum at 
$E_{\rm m} \simeq 2.4 E_{\ast} \simeq 0.15 (\overline{h \nu}/1 \, \rm eV)^{-1} \,
\ \rm TeV$, and then at $E \gg E_{\rm m}$ it decreases as 
$q_{\pm} \propto E^{-(\Gamma+1)} \ln E$. Within an accuracy better than 
20 per cent, the spectrum of secondary pairs  
could be approximated in a simple analytical 
form (Aharonian \& Atoyan 1991):
\begin{equation}
q_{\pm}(E)\ {\rm d}E=f(\Gamma)\frac{\exp{[-(1/(x-1)]}}
{E_{\ast}  x (1+0.07 \ x^{\Gamma}/\ln x)} \ {\rm d} E \, ,
\end{equation}  
where $x=E/E_{\ast} \geq 1$, 
$f(\Gamma)=(1.11-1.60 \Gamma + 1.17 \Gamma^2)$,
and $\int q_{\pm}(E)\ {\rm d}E=2$ (two electrons per interaction).

The intensive synchrotron losses in  the strong magnetic field
$B \sim 100 \, \rm G$  quickly establish a steady-state spectrum 
of electrons proportional to $E^{-2}$  below $E_{\rm m}$, 
and approximately as $E^{-(\Gamma+2)}$  (if we ignore the weak 
logarithmic term) above $E_{\rm m}$. Correspondingly, the 
synchrotron spectrum of the secondary 
pair-produced  electrons is characterized by a  
smooth transition, through the energy around 
$\epsilon_{\rm b}=120 B_{100} (\overline{h \nu}/1 \, \rm eV)^{-2} \, \rm keV$, 
from $s=1.5$ to $s \approx (\Gamma+3)/2$. 
Assuming, for example, that the TeV $\gamma$-rays 
with a photon index $\Gamma \simeq 1.4$ are absorbed 
in an external infrared photon field  
with characteristic energy in the jet frame 
$\overline{h \nu}=1.5 \delta_{10} \, \rm eV$ (like in Fig.~10),  
one should expect {\em in the observer frame} a hard X-ray component 
of radiation with a SED 
$\nu S_\nu \propto \epsilon^{0.5}$ and  $\nu S_\nu 
\propto \epsilon^{-0.2}$ below  and above 
$\epsilon_{\rm b} \approx 500 B_{100} \ \delta_{10}^{-1} \, \rm keV$,
respectively. 
Finally, at energies above several MeV 
the spectrum becomes very steep  
due to the cutoff in the spectrum
of primary (proton-synchrotron) $\gamma$-rays. 
We may speculate that such a component of hard X-rays 
was observed during the strong TeV flares of Markarian 501, 
in particular  in April 1997 (Pian et al. 1998).
Obviously, this component of radiation should 
strongly correlate with the flux of
TeV emission. However, because the   
synchrotron radiation  produced by directly 
accelerated electrons  significantly contributes  to the 
observed  X-ray flux as well, the TeV/X-ray correlation 
may have a rather complicated and non-standard behavior. 
The absence (or suppression)  of such a hard X-ray component in the 
spectrum of Markarian 421 could be explained, within 
the framework of this model, by the  
steep spectrum of TeV $\gamma$-rays  caused by an early synchrotron cut-off. 
The quantitative  study of this question within a detailed time-dependent 
treatment of the problem  will be  discussed elsewhere.

In BL Lac objects like  Markarian 501, 
the photo-meson processes do not play, most probably,  
dominant role in the production
of high energy $\gamma$-rays.
Indeed, the TeV $\gamma$-ray transparency condition  
puts a robust lower limit on the characteristic
time of this process, $t_{\rm p \gamma} \sim 10^{7} \, \rm s$ 
which is almost 3 orders of magnitude larger than the  characteristic 
synchrotron cooling time  of protons.  The severe synchrotron  losses 
of protons cannot be  avoided since the requirement of extremely high 
energy protons in the  blob - a {\em key} assumption in the PIC 
model  -  automatically implies a large 
magnetic field of an order of 100 G.  
On the other hand, the  characteristic photo-meson cooling 
time cannot be arbitrary   reduced,
e.g.  by assuming extremely high density of  radiation in the 
blob.  For any reasonable spectrum of the latter,  and any reasonable geometry 
of the $\gamma$-ray production region,  this would lead to an unacceptably 
large optical depth for TeV $\gamma$-rays. Therefore,  one may conclude that
the very fact of observations of TeV $\gamma$-rays from Markarian 501 and 
Markarian 421 (in fact,  from {\em any} TeV blazar) 
almost rules out the ``photomeson'' 
origin of the bulk of the observed TeV radiation. 
This also implies very low TeV  neutrino fluxes.  

The hypothesis of proton-synchrotron  origin of TeV $\gamma$-flares from 
Markarian 421 and Markarian 501
requires a  very large amount of energy contained in the form of 
magnetic field
\begin{equation}
W_{\rm B}\approx \frac{1}{6}   R^3 \ B^2 =
5.6 \times 10^{49} \ \Delta t_{\rm 3h}^3 \delta_{10}^3  
\ B_{100}^2 \  \rm erg \, .  
\end{equation}
For comparison,  the energy of the magnetic field 
allowed  by the   SSC models ($B \sim 0.1 \, \rm G$) 
is less by six orders of magnitude.
On the other hand, the estimates of the kinetic energy 
in $\geq 10^{19} \, \rm eV$  protons (within the 
proton-synchrotron radiation model) and 
$\geq 10^{12} \, \rm eV$ electrons 
(within the SSC model)  are comparable, if we assume that 
in both models the $\gamma$-rays are  produced with high efficiency.  
Indeed,  in this case the inverse Compton  cooling time of electrons 
and the synchrotron cooling time of protons   
responsible for TeV emission in the SSC 
and proton-synchrotron  models, respectively,  
are  equal or less than the observed variability time-scale
$t^{\ast} \simeq 1.08 \times 10^5 \ \Delta t_{\rm 3h} \delta_{10} \ \rm s$, 
thus 
\begin{equation}
W^{\rm (SSC)}_{\rm e} \sim  W^{\rm (PSR)}_{\rm p} \leq  4  \pi d^2 \
f_\gamma \delta_{\rm j}^{-4} \ t^{\ast} \simeq 1.4 \times 10^{46} \
\Delta t_{\rm 3h} \delta_{10}^{-3} \ \rm  erg \,  .
\end{equation}  
In this estimate the average flux of Markarian 501 in 1997, 
corrected for extragalactic extinction for  ``low-CBR'' model is assumed.
Higher CBR fluxes, as well as the   
possible internal absorption of $\gamma$-rays, 
would increase this estimate by a factor of 3 or so.    
If the spectrum of accelerated particles extends, 
e.g. as $E^{-2}$, down  to energies 
$E \sim m  c^2$, the above estimate 
could be increased by an order of magnitude. In fact,
this simplified estimate only reflects the 
average level of the energy content of 
accelerated particles during the high state of Markarian 501 in 1997.
For the  strongest  flares with duration $\leq 1 \, \rm day$,  the energy
in accelerated protons could be several times larger 
than it follows from  Eq.~(25).  
But in any case,  these uncertainties do not prevent us to conclude 
that in the proton-synchrotron  model we deal with 
a highly magnetized condensation  
of $\gamma$-ray emitting clouds of  EHE protons, where the magnetic pressure 
dominates over  the pressure of relativistic 
protons\footnote{The minimum energy budget condition is 
achieved in the case of approximate 
equipartition between the 
protons and magnetic field, $ W_{\rm B} \approx W_{\rm p}$.
However, for  Markarian 501 
this condition would imply 
magnetic field of about $10 \, \rm G$,  which  
only marginally could provide effective production
of  the proton-synchrotron radiation in the TeV regime.}.
The total magnetic  energy given by Eq.~(24) 
of a single  blob formally is sufficient to supply the  necessary 
energy in  TeV $\gamma$-rays  released 
during  the entire high activity state of Markarian 501 in 1997. 
Thus speculating that an effective 
acceleration  mechanism  operates  in such magnetized condensations,   
stimulated presumably by   
interactions of the latter with the surrounding plasma,  
a single (or a few) energetic blob(s)   
ejected from the central source towards the observer could,
in principle, explain the  extraordinary high TeV 
flux of Markarian 501 in 1997.

In the SSC  models the situation is exactly opposite - the 
pressure of relativistic particles is significantly larger than 
the  pressure of the magnetic field. Even so, 
the total energy contained in the relativistic electrons 
is sufficient to support the 
observed X-ray and $\gamma$-ray emission during 
only several hours, therefore this model 
requires quasi-continuous injection of energy into a single blob, or,
most likely,  a  ``multi-blob'' scenario in which the observed 
radiation is a result 
of superposition  of many  short-live blobs continuously 
ejected from  the central source. 
Then, the  high state of the source 
like the extraordinary long  outburst of Markarian 501 in 1997  
could be associated with a dramatic increase of the  
rate of ejection/formation of $\gamma$-ray emitting blobs.
 
\ack
I  thank the anonymous referee for her/his 
valuable comments, as well as 
S. Wagner, M. Sikora, M. Ostrowski, H. Krawczynski,  
P. Coppi, A. Atoyan, L. Drury, and S. G. Rowell  
for fruitful discussions.

{}
\end{document}